\documentclass{aa}
\usepackage[varg]{txfonts}
\usepackage{longtable}
\usepackage{lscape}
\usepackage{graphicx}
\newcommand{\conum} {\mbox{\rm C/O}}
\newcommand{\kmprs}  {\mbox{\rm km\,s$^{-1}$}}
\newcommand{\feh} {\mbox{\rm [Fe/H]}}
\newcommand{\fehi} {\mbox{\rm [Fe/H]}$_{\rm initial}$}

\newcommand{\xh} {\mbox{\rm [X/H]}}

\newcommand{\oh} {\mbox{\rm [O/H]}}
\newcommand{\ohxx} {\mbox{\rm [O/H]$_{\rm 6300}$}}
\newcommand{\ohyy} {\mbox{\rm [O/H]$_{\rm 7774}$}}

\newcommand{\xfe} {\mbox{\rm [X/Fe]}}
\newcommand{\cfe} {\mbox{\rm [C/Fe]}}
\newcommand{\ofe} {\mbox{\rm [O/Fe]}}
\newcommand{\nafe} {\mbox{\rm [Na/Fe]}}
\newcommand{\mgfe} {\mbox{\rm [Mg/Fe]}}

\newcommand{\alfe} {\mbox{\rm [Al/Fe]}}

\newcommand{\sife} {\mbox{\rm [Si/Fe]}}

\newcommand{\cafe} {\mbox{\rm [Ca/Fe]}}

\newcommand{\tife} {\mbox{\rm [Ti/Fe]}}
\newcommand{\crfe} {\mbox{\rm [Cr/Fe]}}

\newcommand{\nife} {\mbox{\rm [Ni/Fe]}}

\newcommand{\yfe} {\mbox{\rm [Y/Fe]}}

\newcommand{\srfe} {\mbox{\rm [Sr/Fe]}}
\newcommand{\srmg} {\mbox{\rm [Sr/Mg]}}

\newcommand{\ymg} {\mbox{\rm [Y/Mg]}}
\newcommand{\yal} {\mbox{\rm [Y/Al]}}

\newcommand{\alphafe} {\mbox{\rm [$\alpha$/Fe]}}
\newcommand{\teff}  {\mbox{$T_{\rm eff}$}}
\newcommand{\Tc}  {\mbox{$T_{\rm c}$}}

\newcommand{\logg}  {\mbox{{\rm log}\,$g$}}
\newcommand{\logL}  {\mbox{{\rm log}\,$L$}}
\newcommand{\logLsun}  {\mbox{{\rm log}\,$(L/L_{\sun})$}}

\newcommand{\turb}  {\mbox{$\xi_{\rm turb}$}}

\newcommand{\CI} {\ion{C}{i}}
\newcommand{\OI} {\ion{O}{i}}
\newcommand{\oI} {\mbox{\rm [{\ion{O}{i}}]}}

\newcommand{\MgI} {\ion{Mg}{i}}

\newcommand{\SiI} {\ion{Si}{i}}

\newcommand{\CaII} {\ion{Ca}{ii}}
\newcommand{\TiI} {\ion{Ti}{i}}
\newcommand{\TiII} {\ion{Ti}{ii}}
\newcommand{\CrI} {\ion{Cr}{i}}
\newcommand{\CrII} {\ion{Cr}{ii}}

\newcommand{\FeI} {\ion{Fe}{i}}
\newcommand{\FeII} {\ion{Fe}{ii}}
\newcommand{\NiI} {\ion{Ni}{i}}

\newcommand{\YII} {\ion{Y}{ii}}
\newcommand{\SrI} {\ion{Sr}{i}}

\def\ltsima{$\; \buildrel < \over \sim \;$}
\def\simlt{\lower.5ex\hbox{\ltsima}}
\def\gtsima{$\; \buildrel > \over \sim \;$}
\def\simgt{\lower.5ex\hbox{\gtsima}}
\begin{document}

\title{High-precision abundances of elements in solar-type stars}

\subtitle{Evidence of two distinct sequences in abundance-age relations 
\thanks{Tables 1 and  2 are available in electronic form at the CDS via anonymous ftp
to cdsarc.u-strasbg.fr (130.79.128.5) or via http://cdsweb.u-strasbg.fr/cgi-bin/qcat?J/A+A/}}

\author{P. E. Nissen \and J. Christensen-Dalsgaard \and J. R. Mosumgaard \and V. Silva Aguirre \and E. Spitoni \and K. Verma}

\institute{Stellar Astrophysics Centre, 
Department of Physics and Astronomy, Aarhus University, Ny Munkegade 120, DK--8000
Aarhus C, Denmark.  \email{pen@phys.au.dk}.}

\date{Received April 30 2020 / Accepted June 9 2020}

\abstract
{}
{Previous high-precision studies of abundances of elements in solar twin stars are extended
to a wider metallicity range to see how the trends of element ratios with stellar
age depend on \feh .}
{HARPS spectra with signal-to-noise ratios $S/N \simgt 600$ at $\lambda \sim 6000$\,\AA\
were analysed with MARCS model atmospheres
to obtain 1D LTE abundances of C, O, Na, Mg, Al, Si, Ca, Ti, Cr, Fe, Ni, Sr, and Y 
for 72 nearby solar-type stars with metallicities 
in the range $-0.3 \simlt \feh \simlt +0.3$ and ASTEC stellar models were used
to determine stellar ages from effective temperatures, luminosities obtained
via $\it Gaia$ DR2 parallaxes, and heavy element abundances.}
{The age-metallicity distribution appears to consist of two distinct populations: a sequence of old stars
with a steep rise of \feh\ to $\sim \! +0.3$\,dex at an age of $\sim \! 7$\,Gyr and a younger sequence
with \feh\ increasing from about $-0.3$\,dex to $\sim \! +0.2$\,dex
over the last six Gyr. Furthermore, the trends of several 
abundance ratios, [O/Fe], [Na/Fe], [Ca/Fe], and [Ni/Fe],
as a function of stellar age split into
two corresponding sequences. The [Y/Mg]-age relation, on the other hand,
shows no offset between the two age sequences and has no significant dependence on [Fe/H], 
but the components of a visual binary star, $\zeta$\,Reticuli, have a 
large and puzzling deviation.}
{The split of the age-metallicity distribution into two sequences
may be interpreted as evidence of two episodes of accretion of gas onto 
the Galactic disk with a quenching of star formation in between. 
Some of the [X/Fe]-age relations support this scenario but other relations are 
not so easy to explain, which calls for a deeper study of systematic errors
in the derived abundances as a function of [Fe/H], in particular 3D non-LTE effects.}


\maketitle

\section{Introduction}
\label{introduction} 
The relation between ages of stars and their metallicities is of fundamental
importance for studies of the chemical evolution of our galaxy. According to the simple closed
box model with instantaneous mixing of produced elements \citep{pagel97},
one expects \feh\,\footnote{For two elements, X and Y,
with number densities $N_{\rm X}$ and $N_{\rm Y}$,
[X/Y] $\equiv {\rm log}(N_{\rm X}/N_{\rm Y})_{\rm star}\,\, - 
\,\,{\rm log}(N_{\rm X}/N_{\rm Y})_{\sun}$.}  
to increase smoothly with time. The spectroscopic
survey of \citet{edvardsson93} showed, however, that the mean metallicity of
disk stars in the solar neighbourhood is nearly constant for ages between
2 and 10\,Gyr but with a large scatter in \feh\ at a given age corresponding to a range from 
$-0.4$ to +0.3\,dex. This has been confirmed in many later
investigations using different techniques to derive stellar metallicities and
ages such as Str{\"o}mgren photometry \citep{Feltzing01, nordstrom04,
casagrande11}, spectroscopy and luminosities based on {\it Hipparcos} or {\it Gaia} distances
\citep[e.g.][]{bensby14, buder19, delgado-mena19}, and APOGEE abundances combined with
asteroseismic ages \citep{silva-aguirre18, miglio20}. 

\citet{edvardsson93} suggested that the dispersion in the age-metallicity relation
could be due to infall of metal-poor gas onto the Galactic disk triggering star formation.
The canonical explanation is, however, that the dispersion is due to radial migration 
of stars in a disk with a gradient in \feh\
\citep[e.g.][]{sellwood02, minchev13, frankel18, feuillet19}.                     
Given that the radial gradient of \feh\ is around $-0.06$\,dex\,kpc$^{-1}$
\citep{anders17}, mixing
must then take place over several kpc. This cannot be explained by increasing amplitudes
in the epicycle motion of stars (`blurring'), but requires changes of
the orbital angular momentum (`churning') for example due to perturbations from spiral
waves in the disk \citep{schonrich09}.

In contrast to the scatter in the age-metallicity relation, there is a tight
correlation between age and \alphafe\,\footnote{$\alpha$ denotes the abundance of 
alpha-capture elements; in this paper defined as the mean abundance of Mg, Si, and Ti.} 
for stars in the solar neighbourhood
\citep{haywood13, bensby14, delgado-mena19}.
For the oldest stars, \alphafe\ declines steeply with decreasing age, but from an age
of $\sim \! 7$\,Gyr the decrease is only $\sim \! 0.05$\,dex until the present time.
Furthermore, high-precision abundance studies of solar twins \citep{dasilva12, nissen15, spina16,
bedell18} have revealed the existence of tight age correlations for other element ratios.
In particular, there is a large variation in \ymg\ from about $-0.20$\,dex
at an age of 10\,Gyr to +0.15\,dex for the youngest stars 
\citep{nissen16, tucci-maia16, spina18}
suggesting that \ymg\ can be used as a sensitive chemical clock to obtain stellar ages. 
Because the solar twins are confined to a small metallicity range,
$-0.1 < \feh < +0.1$, it is, however, unclear if the \ymg -age relation is 
valid at other metallicities. \citet{feltzing17} and \citet{delgado-mena19}
find a significant shift of the \ymg -age relation with \feh , but this is not confirmed by
\citet{titarenko19}.

In order to see if the age relations of \ymg\ and other element ratios depend on metallicity 
we have expanded the high-precision study of solar twins by 
\citet{nissen15, nissen16} and \citet{nissen17} to solar-type  stars 
covering the metallicity range from $-0.3$ to +0.3\,dex. Interestingly, this has
provided evidence of two distinct populations in the age-metallicity 
diagram as well as corresponding sequences in the trends of 
element ratios with stellar age. These bimodal structures have not been seen before,
but are probably related to the high- and low-alpha sequences in 
the  \alphafe -\feh\ diagram known for stars in the solar vicinity
\citep[e.g.][]{fuhrmann98, gratton00, prochaska00, bensby05, reddy06, adibekyan12}
and also at distances of several kpc \citep[e.g.][]{nidever14, hayden15}.

The stellar atmospheric parameters and chemical abundances as derived from a 1D, LTE
model atmosphere analysis of HARPS spectra
are presented in Sect. \ref{abundances}, and the determination of stellar ages
from effective temperatures and luminosities based on $Gaia$ DR2 parallaxes 
in Sect. \ref{ages}. The resulting age - abundance trends are presented in 
Sect. \ref{results} including a discussion of potential systematic errors due to 3D non-LTE
effects and a discussion of the binary star $\zeta$\,Reticuli. 
Possible explanations of the two distinct sequences in the age-metallicity
diagram are discussed in Sect. \ref{discussion} and a summary with some conclusions 
is given in Sect. \ref{conclusions}. 

\section{Stellar parameters and elemental abundances}
\label{abundances}
Based on the effective temperatures, surface gravities and metallicities derived 
by \citet{sousa08} for stars in the HARPS-GTO planet search program \citep{mayor03}, 
we first selected stars that are not spectroscopic binaries and have
5600\,K $< \teff < 5950$\,K, $\logg > 4.15$, and $-0.3 < \feh < +0.3$.
After combining the HARPS spectra in the ESO Science Archive,
the signal-to-noise (S/N) was checked and a sub-sample of stars having spectra 
with $S/N > 600$ at $\lambda \sim 6000$\,\AA\ were
selected so that their distribution in metallicity is approximately uniform. 
The sample includes the solar twins analysed in 
\citet[][hereafter Papers\,I and II]{nissen15, nissen16} 
and the visual binary star 16\,Cyg\,A and B (HD\,186408 and HD186427),
observed with the HARPS-N instrument at the TNG 3.5\,m telescope 
and analysed in \citet[][Paper\,III]{nissen17}. Furthermore, six young solar twins 
(HD\,6204, HD\,12264, HD\,59967, HD\,75302, HD\,196390, and HD\,202628)
were included from \citet{bedell18}. Such stars were avoided in the HARPS-GTO
program, because their high magnetic activity was judged to make detection of planets difficult.

Details about the normalisation of HARPS spectra and the measurements of equivalent
widths (EWs) of spectral lines may be found in Paper\,I and a more general discussion
of methods in high-precision abundance studies are given by \citet{nissen18}.
Here, we stress the importance of making a differential analysis relative
to the Sun represented by a $S/N \simeq 1200$ HARPS solar flux spectrum observed
via reflected sunlight from the minor planet Vesta. Furthermore, it is important that
the stars have been selected as main-sequence stars ($\logg > 4.15$) having
effective temperatures within $\pm 180$\,K
from the temperature of the Sun (assumed to be $\teff = 5777$\,K). This ensures that
the strengths of spectral lines are not very different and that the same continuum 
windows can be used when measuring EWs.

Abundances were derived from the list of spectral lines given in Table\,2 of Paper\,I.
For stars with $\feh < -0.15$ and $\feh > +0.15$, 13 of the 132 lines used for the
solar twins gave abundances that deviated in a systematic way from the mean abundances
of a given element suggesting that the measured EWs 
of these lines\,\footnote{The excluded lines are:
\MgI\ 4730.04\,\AA , \SiI\ 5793.08\,\AA , \TiI\ 5739.48 and 5866.46\,\AA ,
\TiII\ 5381.03\,\AA , \CrI\ 5247.57, 5296.70, and 5348.33\,\AA ,
\FeI\ 5466.99 and 6157.73\,\AA , \FeII\ 6416.93\,\AA , \NiI\ 6108.12 and
6643.64\,\AA .} are affected by blends or weak lines in the continuum regions applied.
They were therefore excluded when calculating
the final mean abundances of the elements. In the case of Mg, there is then only one line
left, i.e. the \MgI\ 5711.10\,\AA\ line; it is judged to provide a more reliable
abundance than the \MgI\ 4730.04\,\AA\ line, which occurs in a crowded wavelength
region. 

Sulphur and zinc, which are represented by respectively four and three lines in
Table\,2 of Paper\,I, are not included in this paper, because the line-to-line scatter
of the derived abundances for stars with $\feh < -0.15$ and $\feh > +0.15$
is significantly higher than expected and it is unclear which lines are problematic. 
On the other hand, one new line was included, the strontium \SrI\ line at 4607.34\,\AA ,
to have another $s$-process element represented in addition to yttrium.

Assuming local thermodynamic equilibrium (LTE), the Uppsala EQWIDTH program 
was used to calculate equivalent widths as a function of element abundance
for model atmospheres obtained by interpolation in the 1D MARCS grid 
\citep{gustafsson08}\,\footnote{{\tt https://marcs.astro.uu.se}} to
the \teff , \logg , \feh , and \alphafe\ values of the stars.
The observed EWs then provide abundances
for each spectral line. Of particular importance are the iron abundances,
because they are used to determine the atmospheric parameters   
\teff , \logg , and microturbulence (\turb ) by requesting that \feh\ has no systematic 
dependence on excitation potential, ionisation stage and EW of the lines. 
\alphafe\ was included as a parameter of the models, because it affects the electron 
pressure and therefore the determination of \logg . Altogether, this means that the determination
of abundances and parameters is an iterative process, which was continued until
the change of \teff\ was less than 2\,K, change of \turb\ less than 0.01\,\kmprs ,
and the changes of \logg , \feh , and \alphafe\ less than 0.002\,dex.

\begin{table*}
\caption[ ]{Stellar atmospheric parameters, heavy element abundance, luminosity, age, mass, photometric gravity, and
helium abundance. The full table is available at the CDS.}
\label{table:param}
\centering
\setlength{\tabcolsep}{0.20cm}
\begin{tabular}{rcccrrcrccccc}
\noalign{\smallskip}
\hline\hline
\noalign{\smallskip}
\noalign{\smallskip}
HD no. & \teff & \logg  & \turb &  \feh & \alphafe & $Z_s$ & log$(L / L_{\sun}$) & Age & $\sigma (Age)$ &
$M / M_{\sun}$ & \logg & $Y_s$ \\
       & [K]   & (spec) &  \kmprs &    &          &       &          &[Gyr]& [Gyr]          &
                  & (phot)&       \\
\noalign{\smallskip}
\hline
\noalign{\smallskip}
     361 &  5892 &   4.524 &   1.05 &  $-$0.130 &  $-$0.003 &   0.0135 &  $-$0.021 &     1.5 &   0.9 &    1.01 &   4.497 &   0.255\\
    1461 &  5760 &   4.372 &   1.03 &   0.190 &   0.009 &   0.0275 &   0.077 &     5.5 &   0.5 &    1.05 &   4.378 &   0.257\\
    2071 &  5724 &   4.486 &   0.96 &  $-$0.087 &   0.013 &   0.0152 &  $-$0.076 &     4.1 &   0.9 &    0.96 &   4.481 &   0.245\\
     ... &   ... &     ... &    ... &     ... &     ... &      ... &     ... &     ... &  ...  &     ... &     ... &     ...\\
  211415 &  5853 &   4.381 &   1.11 &  $-$0.251 &   0.037 &   0.0110 &   0.066 &     7.7 &   0.7 &    0.94 &   4.369 &   0.220\\
  220507 &  5692 &   4.238 &   1.07 &   0.004 &   0.101 &   0.0215 &   0.165 &     9.7 &   0.5 &    1.00 &   4.248 &   0.235\\
  222582 &  5784 &   4.359 &   1.07 &  $-$0.014 &   0.031 &   0.0184 &   0.096 &     7.4 &   0.5 &    1.00 &   4.344 &   0.234\\
\noalign{\smallskip}
\hline
\end{tabular}
\end{table*}

\begin{table*}
\caption[ ]{Stellar abundance ratios. The full table is available at the CDS.}
\label{table:abun}
\centering
\setlength{\tabcolsep}{0.16cm}
\begin{tabular}{rrrrrrrrrrrrr}
\noalign{\smallskip}
\hline\hline
\noalign{\smallskip}
\noalign{\smallskip}
  HD no. & \cfe    &  \ofe   & \nafe  &   \mgfe &   \alfe &   \sife  &   \cafe &   \tife &  \crfe  &   \nife &   \srfe &   \yfe \\
\noalign{\smallskip}
\hline
\noalign{\smallskip}
     361 &  $-$0.039 &   0.010 &  $-$0.072 &  $-$0.014 &  $-$0.042 &  $-$0.013 &   0.027 &   0.019 &   0.002 &  $-$0.056 &   0.106 &   0.079\\
    1461 &  $-$0.032 &  $-$0.108 &   0.103 &   0.007 &   0.044 &   0.014 &  $-$0.016 &   0.007 &   0.005 &   0.045 &  $-$0.052 &  $-$0.039\\
    2071 &  $-$0.020 &  $-$0.022 &  $-$0.032 &   0.001 &  $-$0.001 &   0.002 &   0.024 &   0.029 &   0.005 &  $-$0.029 &   0.032 &   0.039\\
     ... &     ... &     ... &    ... &     ... &     ... &      ... &     ... &     ... &    ...  &     ... &     ... &     ...\\
  211415 &   0.030 &   0.057 &   0.001 &   0.048 &   0.030 &   0.032 &   0.028 &   0.032 &  $-$0.012 &  $-$0.019 &  $-$0.009 &  $-$0.090\\
  220507 &   0.097 &   0.153 &   0.040 &   0.125 &   0.153 &   0.069 &   0.050 &   0.109 &   0.004 &   0.016 &  $-$0.087 &  $-$0.085\\
  222582 &   0.001 &   0.042 &   0.018 &   0.035 &   0.059 &   0.028 &   0.015 &   0.030 &   0.000 &   0.005 &  -0.051 &  -0.066\\
\noalign{\smallskip}
\hline
\end{tabular}
\end{table*}

In Papers\,I, II, and III, abundances were derived 
under the assumption that all stars have the same atmospheric helium-to-hydrogen ratio 
as the Sun, i.e.  $N_{\rm He} / N_{\rm H} = 0.085$, corresponding to a 
helium mass fraction of $Y = 0.249$.  There are, however, differences in the 
helium abundances due to Galactic chemical evolution and
He diffusion in stars \citep[see e.g.][]{verma19}.
These variations affect the hydrogen abundance and therefore the derived \xh\ values.
We take this into account by using the helium abundances
calculated in connection with the stellar age determinations (see Table \ref{table:param}).  
The effect on \xh\ for the solar twins in Paper\,I are within $\pm 0.01$\,dex,
but reach about $-0.02$\,dex for the oldest, most metal-poor stars with $Y \simeq 0.21$
and +0.02\,dex for the youngest metal-rich stars with $Y \simeq 0.28$. 
The changes in \xfe\ are, on the other hand, almost negligible,
because the change of abundance with respect
to hydrogen is nearly the same for all elements except in cases where the abundance
is derived from a line with damping wings such as the \MgI\ 5711.10\,\AA\ line.
For this line, the $N_{\rm He} / N_{\rm H}$ ratio has a small effect 
($< 0.01$\,dex) on the derived \mgfe\ ratio,
because H and He atoms contribute to the van der Waals broadening with different amounts
\citep[][p. 217]{gray92}.

In addition to the direct effect of a change in He abundance on the derived \xh\ values,
there is also an effect on the pressure structure
of the model atmospheres, which was not taken into account, because the MARCS models
are only available for $N_{\rm He} / N_{\rm H} = 0.085$ (the solar value).
This has small effects on some abundance ratios and also on the gravities 
obtained by comparing iron abundances derived from \FeI\ and \FeII\ lines
as discussed is Sect. \ref{ages}.

The derived atmospheric parameters and \xfe\ abundance ratios are given in 
Tables \ref{table:param} and \ref{table:abun}.
Small non-LTE corrections were included in the published abundances for the solar twins
in Papers\,I, II, and III, but because 3D effects may be as important as non-LTE
effects, we prefer here to give the 1D, LTE results. 
Full 3D, non-LTE corrections may then be applied when they become available.
Such corrections are already available for C and O lines 
\citep{amarsi19b} and will be discussed in Sect. \ref{results}.

The statistical errors of the atmospheric parameters 
were estimated with the error analysis method of \citet{epstein10} and
errors of abundance ratios were calculated as the quadratic sum of
errors arising from the uncertainty of the atmospheric parameters and the
EW measurements. The latter contribution was calculated from the line-to-line
scatter of \xfe\ for groups of stars with similar spectra using Eqs. 2 - 4 in
Paper\,II. For elements with only one line (O, Mg, and Sr), the error of the EW was
estimated from the scatter of repeated measurements. As stars with small or large values
of \feh\ have the most deviating spectra from that of the Sun, 
we have divided the sample into three groups with respectively
$\feh < -0.10$, $-0.10 \le \feh \le +0.10$ (solar twins), and $\feh > +0.10$. 
The average errors for these groups are given in Table \ref{table:errors}.

\begin{table}
\caption[ ]{Average statistical (1-sigma) errors for three groups of stars}
\label{table:errors}
\centering
\setlength{\tabcolsep}{0.15cm}
\begin{tabular}{rccc}
\noalign{\smallskip}
\hline\hline
\noalign{\smallskip}
        & $\feh < -0.10$ & Solar twins & $\feh > +0.10$ \\
\noalign{\smallskip}
        & $N_{\rm star} = 28$  & $N_{\rm star} = 27$ & $N_{\rm star} = 17$   \\
\noalign{\smallskip}
\hline
\noalign{\smallskip}
$\sigma$\teff & 9\,K  &  6\,K & 10\,K \\
$\sigma$\logg & 0.018 & 0.012 & 0.020 \\
$\sigma$\turb & 0.03 \kmprs & 0.02 \kmprs & 0.03 \kmprs \\
$\sigma$\feh  & 0.009 & 0.006 & 0.010 \\
$\sigma$\alphafe & 0.006 & 0.005 & 0.006 \\
$\sigma$\cfe  & 0.016 & 0.013 &  0.024 \\
$\sigma$\ofe  & 0.030 & 0.022 &  0.030 \\
$\sigma$\nafe & 0.007 & 0.007 &  0.007 \\
$\sigma$\mgfe & 0.014 & 0.011 &  0.014 \\
$\sigma$\alfe & 0.008 & 0.008 &  0.010 \\
$\sigma$\sife & 0.007 & 0.007 &  0.008 \\
$\sigma$\cafe & 0.006 & 0.006 &  0.006 \\
$\sigma$\tife & 0.007 & 0.007 &  0.009 \\
$\sigma$\crfe & 0.006 & 0.006 &  0.007 \\
$\sigma$\nife & 0.006 & 0.006 &  0.006 \\
$\sigma$\srfe & 0.016 & 0.012 &  0.027 \\
$\sigma$\srmg & 0.025 & 0.017 &  0.033 \\
$\sigma$\yfe  & 0.013 & 0.010 &  0.013 \\
$\sigma$\ymg  & 0.020 & 0.014 &  0.020 \\
\noalign{\smallskip}
\hline
\end{tabular}
\end{table}

\section{Stellar ages}
\label{ages}
The age of a star was determined from its \teff , luminosity $L$,
and atmospheric (surface) heavy element mass fraction $Z_s$ by interpolating 
in a grid of stellar models 
computed with the  Aarhus STellar Evolution Code \citep[ASTEC, see][]{jcd08a}.
These models include diffusion and settling of helium and heavier elements according
to \citet{michaud93}, which means that the
surface abundances are reduced approximately proportional to
age with about 2\% per Gyr.
The models range in mass from 0.7 to 1.2\,$M_{\sun}$ and in initial 
heavy element mass fraction from $Z_i = 0.008$ to  $Z_i = 0.035$
on a scale where the solar heavy element mass fraction is $Z_{\sun} = 0.0180$  
as calculated from the \citet{grevesse96} solar atmospheric composition. 
The mixing length parameter ($\alpha_{\rm ML} = l/H_p$) is assumed to be independent 
of mass, age, and metallicity with a value of $\alpha_{\rm ML} = 2.08$ obtained by fitting
the model of the Sun to the solar radius and luminosity at an age of 4.6\,Gyr
and a solar surface helium abundance of $Y_{\sun} = 0.245$ 
\citep[close to the value determined from helioseismology, e.g.][]{jcd91, vorontsov91, basu04}.
The initial helium mass fraction is assumed to  
vary as $\Delta Y_i / \Delta Z_i = 1.4$, which ratio is well within the range,
$\Delta Y_i / \Delta Z_i = 1.23 \pm 0.85$, determined by \citet{verma19} from 
a study of acoustic glitches in stellar oscillation frequencies of 38 stars
in the {\em Kepler} LEGACY sample.

The V-band luminosities of the stars were calculated from Johnson $V$ magnitudes  
\citep{olsen83} and distances based on $Gaia$ DR2 parallaxes \citep{gaia.collaboration18}
except in the case of HD\,210918 for which we adopted the 
Hipparcos parallax \citep{vanleeuwen07},
because this star is not in the $Gaia$ catalogue. As all stars are closer than 60\,pc, we assumed that 
they are not affected by interstellar absorption and reddening, which is supported by
colour excesses, $E(b-y)$, calculated from Str{\"o}mgren $uvby-H_{\beta}$ photometry of
\citet{olsen83} and the $(b-y)_0 - \beta$ calibration of \citet{schuster89}; 
the sample has an average $E(b-y)$ of 0.002\,mag with an rms scatter of 0.009\,mag only.

Bolometric corrections were estimated from the \citet{casagrande10} $V\! - \!K$ calibration 
with $K$ magnitudes taken from the 2MASS catalogue of \citet{cutri03}. 
As the stars have effective temperatures and metallicities not too different
from the solar values, the $BC$ corrections differ by less than
$\pm 0.04$\,mag from the bolometric correction, $-0.07$\,mag, for the Sun,
for which a bolometric magnitude of $M_{\rm bol} = -4.74$ was adopted.

In calculating the statistical uncertainty of the derived bolometric magnitudes,
we adopt an error of 0.02\,mag in $V$ and 0.01\,mag in $BC$. To this comes the error 
arising fror the relative error of the parallax, i.e. $\sim 2.17 \sigma(p)/p$,
which range from 0.005 to 0.016\,mag. This transform to errors of 
\logLsun\ ranging from 0.009 to 0.011\,dex. In this connection, we note
that correction of a possible systematic error of the $Gaia$ DR2 parallaxes of $-0.03$\,mas
\citep{lindegren18} affects \logL\ by less than 0.002\,dex. Given that
the size of this systematic offset of the $Gaia$ DR2 parallaxes is uncertain, 
it has not been included.

The ASTEC models are all calculated for the \citet{grevesse96} 
solar mixture of elements relative to Fe.
Following  \citet{salaris93}, the enhancement of  $\alpha$-capture elements 
(O, Ne, Na, Mg, Si, P, S, Cl, Ar, Ca, and Ti)
relative to Fe was  taken into account by calculating the
surface heavy element mass fraction as
\begin{eqnarray} 
Z_s \, = \, Z_0  \, (0.694 \, \cdot \, 10^{\alphafe} + 0.306).
\end{eqnarray}
Here the coefficient 0.694 is the fractional contribution to $Z_{\sun}$ from 
$\alpha$-capture elements and $Z_0$ is the $Z$-value that would be calculated 
without enhancement, i.e. from the expression
\begin{eqnarray} 
\frac{Z_0}{X_0} \, = \, \frac{Z_{\sun}}{X_{\sun}} \, 10^{\feh},  
\end{eqnarray}
where the mass fraction of hydrogen is calculated from the relations
$X_0 \, = \, 1 - Y_0 - Z_0$ and $Y_0 \, \simeq \, Y_{\sun} + 1.4 \, (Z_0 - Z_{\sun})$.

Values of \teff , \logLsun , and $Z_s$ are given in Table \ref{table:param} together with
the derived stellar ages, masses and surface helium abundances.
The listed errors of the ages are estimated as a quadratic sum of 
1-sigma errors arising from the uncertainties in \teff , \logL , and $Z_s$. 
They are statistical errors showing how well relative ages at a given \feh\
have been determined. Absolute ages are more uncertain; in particular there 
will be systematic changes of the derived ages as a function of \feh\ if
a different $\Delta Y_i / \Delta Z_i$ ratio is assumed or if the \citet{asplund09}
solar mixture of the elements is adopted instead of the \citet{grevesse96} mixture.
As can be seen from the table, the age errors lie in the range 0.5-1.3\,Gyr with
the largest errors occurring for the cooler stars close to the ZAMS, where the
isochrones are closely packed (see Fig. \ref{fig:iso.zetaRet}). 
In this connection, we note that one star,
HD\,6204, lies slightly below the ZAMS corresponding to its $Z_s$. It is formally
estimated to have a `negative' age of $-0.3 \pm 1.0$\,Gyr, but is plotted with
zero age in the various figures.

As a sanity check of the ASTEC ages, we have 
used a grid of stellar models computed with the Garching Stellar
Evolution Code \citep[GARSTEC,][]{weiss08} to determine ages
from \teff , $L$, \feh , and \alphafe\
with the Bayesian Stellar Algorithm (BASTA) described in \citet{silva-aguirre15}.
The GARSTEC models include diffusion of helium and heavy elements according to the 
\citet{thoul94} prescription and are  based on the $\Delta Y / \Delta Z = 1.4$ assumption
like the ASTEC models.
The grid comprise the same mass range as the ASTEC models but with a smaller step
in initial composition ($\Delta \feh = 0.05$) and with \alphafe\ having
values of 0.0, 0.1, 0.2, and 0.3\,dex relative to a solar composition 
adopted from \citet{asplund09}. Based on this grid, BASTA was used to find the
most likely age and its positive and negative 1-sigma errors assuming as priors 
an age range from 0 to 12\,Gyr,
a standard Salpeter Initial Mass Function, and an interval 
for the mixing length parameter from 1.7 to 1.9 with the solar value
being $\alpha_{\rm ML} = 1.79$.

\begin{figure}
\resizebox{\hsize}{!}{\includegraphics{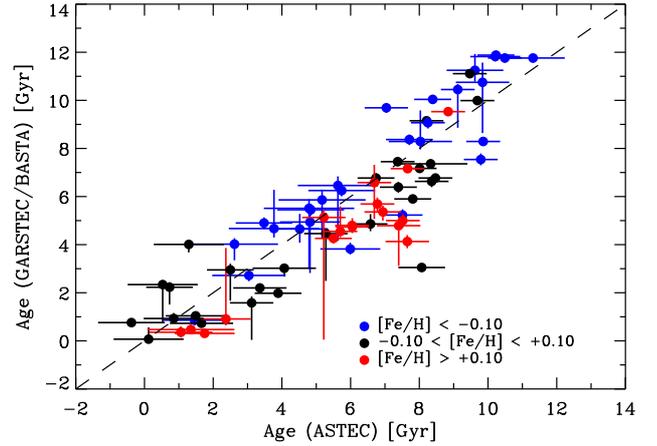}}
\caption{Comparison of stellar ages determined from GARSTEC models with the BASTA program 
and ages determined from ASTEC models. The stars have been divided into three
metallicity groups as indicated.}
\label{fig:ages}
\end{figure}

Figure \ref{fig:ages} shows the comparison between the BASTA and ASTEC ages
for three ranges of \feh . Overall, there is a satisfactory agreement between the
two set of ages although some stars deviate more than expected based on the estimated errors.
Metal-poor stars tend, however, to be a bit older on the  BASTA age scale,
whereas metal-rich stars are on average about 1\,Gyr younger than the ASTEC ages.  
This trend may be related to the difference in the adopted solar composition.
The \citet{asplund09} mixture, used in GARSTEC, has a lower O/Fe  ratio 
than the \citet{grevesse96} mixture adopted in ASTEC,
which means that the GARSTEC models are less sensitive to the variations in \alphafe\
from high values in metal-poor stars to low values in metal-rich stars.

\begin{figure}
\resizebox{\hsize}{!}{\includegraphics{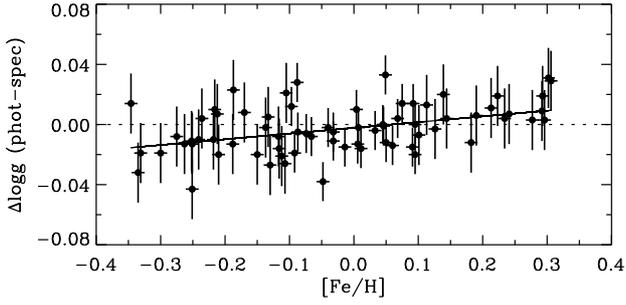}}
\caption{Difference between photometric and spectroscopic gravities 
as a function of \feh . The line shows the fit to the data given by
Eq. (4).}
\label{fig:logg}
\end{figure}

Stellar ages could also have been derived by using 
spectroscopic gravities instead of luminosities as in Papers\,I and II, where
the available Hipparcos parallaxes did not allow a determination of $L$ with sufficient
precision. Conversely, we can make a comparison between luminosities and spectroscopic 
gravities by calculating a `photometric' gravity from the
expression
\begin{eqnarray} 
\logg \,({\rm phot}) = 4.438 + \log \frac{M}{M_{\sun}} +
4 \log \frac{\teff}{T_{\rm eff,\sun}}  - \log \frac{L}{L_{\sun}}.
\end{eqnarray}
The average difference between the photometric gravity and the spectroscopic
gravity is $-0.003$\,dex with an 
rms deviation of 0.016\,dex, which can be explained by the estimated errors
of the spectroscopic gravities. As seen from Fig. \ref{fig:logg}, there is, however, 
a slight trend of the difference with \feh :
\begin{eqnarray}
\Delta \logg = -0.002 \, (\pm 0.002) + 0.038 \, (\pm 0.013) \cdot \feh , 
\end{eqnarray}
as found from a linear regression to the data.

We have first investigated if the trend of $\Delta \logg$ with \feh\
arises because the MARCS models applied for the determination of
spectroscopic gravities  have a constant helium-to-hydrogen ratio 
$y = N_{\rm He}/N_{\rm H} = 0.085$,
whereas the He abundances of the stars depends on \feh\ and age. 
Helium does not contribute to opacity or electrons in solar-type stars
but increases the mean molecular weight by a factor of 1+4$y$ 
and atomic pressure by a factor of 1+$y$ relative to the contributions from hydrogen.
As shown by \citet[][Eq. 12]{stromgren82},
a change in the helium-to-hydrogen ratio from $y_1$
to $y_2$ has the same effect on line strengths as a change in gravity from $g_1$ to $g_2$,
where
\begin{eqnarray}
g_2 \, = \, g_1 \, \frac{1 + 4 y_1}{1 + y_1} \,\,\frac{1 + y_2}{1 + 4 y_2},
\end{eqnarray}
provided that the electron pressure is much smaller than the gas pressure
as in the upper layers of the atmospheres of solar-type stars.
This equation was used to calculate corrected spectroscopic gravities using
$y_1 = 0.085$ and $y_2$ values corresponding to the helium mass fractions in 
Table \ref{table:param}. The correction
goes, however, in the wrong direction and increases the slope of $\Delta \logg$
versus \feh\ by nearly a factor of two. Hence, there must be other systematic
errors in the analysis depending on \feh , most likely 3D non-LTE effects on the
relative strengths of \FeI\ and \FeII\ lines, but deviations from 
assumptions related to the ASTEC models, such as $\Delta Y_i / \Delta Z_i = 1.4$ and a constant
mixing length parameter could also play a role.  

While the trend of $\Delta \logg$\,(phot.\,$-$\,spec.) with \feh\ is an interesting problem,
it has only a small effect on the derived ages and abundances. If
spectroscopic gravities are used in the age determination instead of luminosities, the ages
of young, metal-poor stars change by about $-0.5$\,Gyr and ages of young, metal-rich stars
by $\sim +0.5$\,Gyr. For older more evolved stars, the changes are smaller. Concerning
\xfe\ abundance ratios, the largest effects of using photometric instead of spectroscopic gravities  
are on the order of $\pm 0.005$\,dex in the case of  C, O, Mg, and Y. 
These changes are well within the estimated uncertainty of the abundances 
(see Table\,\ref{table:errors}).

\section{Results}
\label{results}
In this section, we first show that there are indications of two distinct sequences
in the age-metallicity and \xfe -age diagrams. Next we discuss possible
3D, non-LTE effects on the results. Furthermore, we present the \srmg - and \ymg -age relations
and discuss their dependence on metallicity with particular emphasis of a puzzling
deviation of the visual binary star, $\zeta$\,Reticuli.

\begin{figure}
\resizebox{\hsize}{!}{\includegraphics{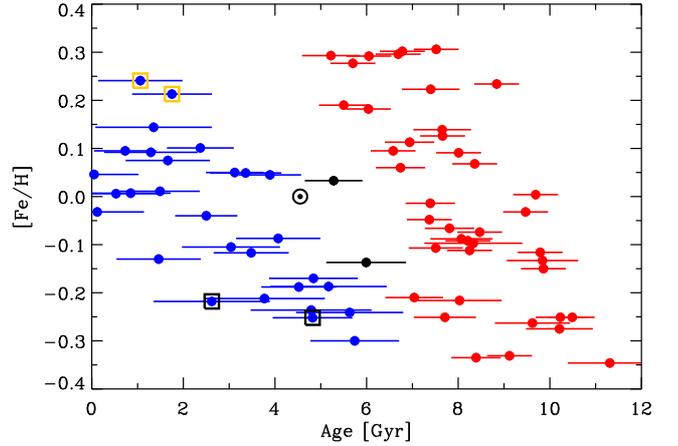}}
\caption{\feh\ versus stellar age. The stars have been divided in two groups:
an old sequence shown with filled red circles and a younger sequence shown with
filled blue circles. Two stars (HD\,59711 and HD\,183658) having intermediate
ages are shown with filled black circles and the Sun with the $\odot$ symbol.
The components of the visual binary star, $\zeta$\,Reticuli, are marked with black squares
and the Na-rich stars, HD\,13724 and HD\,189625, with yellow squares.}
\label{fig:feh-age}
\end{figure}

\begin{figure*}
\resizebox{\hsize}{!}{\includegraphics{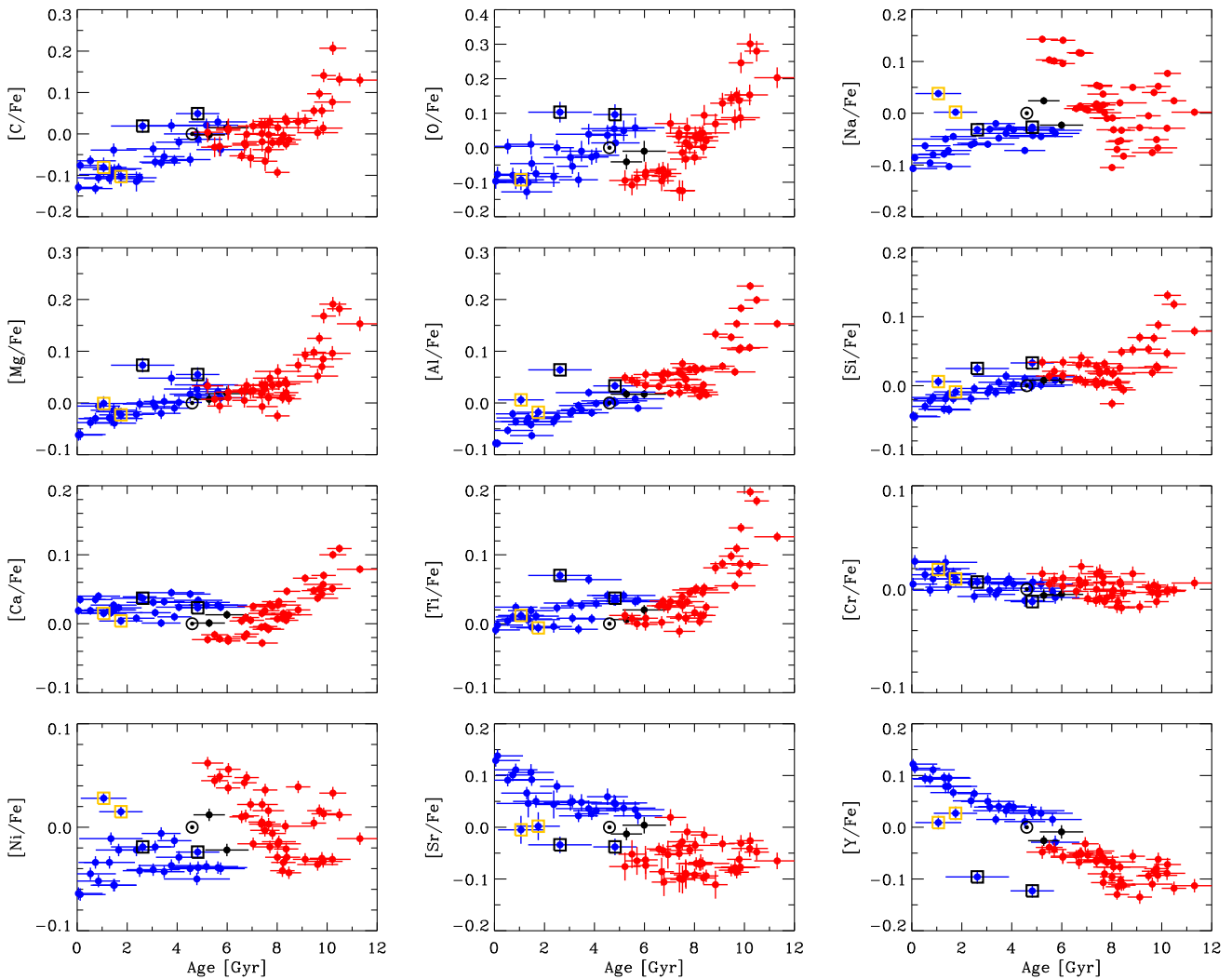}}
\caption{1D LTE values of \xfe\ as a function of stellar age for all elements included.
Stars are shown with the same symbols as in Fig. \ref{fig:feh-age}.}
\label{fig:xfe-age}
\end{figure*}

\subsection{\feh\ and \xfe\ versus stellar age}
\label{abundance-age}
The relation between \feh\ and stellar age is shown in Fig. \ref{fig:feh-age}.
As seen, the stars tend to be distributed in two populations: i.e. an old
sequence (red filled circles) reaching $\feh \sim \, +0.3$ at an age of $\sim 7$\,Gyr
and a younger sequence (blue filled circles) stretching from $\feh \simeq -0.3$ at
6\,Gyr to $\feh \simeq +0.2$ at $\sim 1$\,Gyr. Because stars were
selected to have 5600\,K $< \teff < 5950$\,K, there may be a bias against old metal-rich
stars in the upper right corner of the figure and against young metal-poor stars in
the lower left corner, but we see no selection effects that could explain the 
dearth of stars at intermediate ages. Furthermore, although there may be systematic errors
affecting the relative ages of stars with different metallicities as discussed above, 
these errors do not affect the differential ages of stars at a given \feh .
The sample is, however, small, so it could be accidental that there are so few
stars between the two populations. Clearly, the possible existence of two
distinct sequences in the age-metallicity diagram should be investigated for a larger
sample of stars with precise ages and abundances.

Additional evidence of the existence of two distinct chemical evolution sequences 
comes from the \xfe -age relations shown in Fig. \ref{fig:xfe-age}.
For several of the elements, e.g. O, Na, Ca, and Ni, there is a discontinuity or bump in \xfe\ at 
$\sim 6$\,Gyr between the old (red) and young (blue) sequences.

\begin{figure}
\resizebox{\hsize}{!}{\includegraphics{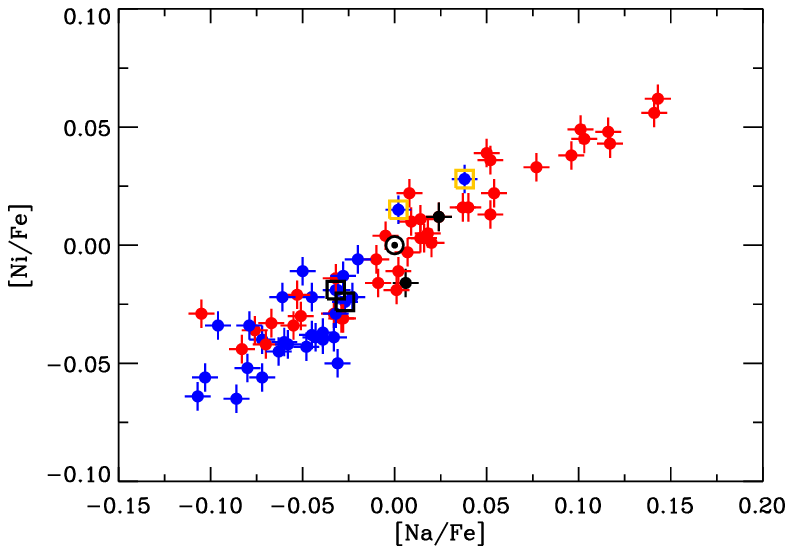}}
\caption{\nife\ versus \nafe . Stars are shown with the same symbols as in Fig. \ref{fig:feh-age}.}
\label{fig:nife-nafe}
\end{figure}

As seen from Fig. \ref{fig:xfe-age},
the components of the binary star $\zeta$\,Reticuli, marked with black squares,
show a strong deviation from the mean trends of the $s$-process elements (Sr and Y)
and they also deviate in \cfe , \ofe , and \mgfe . This binary star will be 
discussed in Sect. \ref{zetaRet}. 
Furthermore, two stars, HD\,13724 and HD\,189625, marked with yellow squares,   
are enhanced in Na and Ni relative to Fe, whereas they are under-abundant
in Sr and Y. Excluding these stars, the scatter in the age trends can be explained by 
the estimated error bars except in the case of the old sequences for Na and Ni. 
For these two elements, there is on the other hand a very good correlation
between \nife\ and \nafe\ (see Fig. \ref{fig:nife-nafe}) indicating that the scatter
in the age relations are not due to errors in the abundance determinations.
A tight correlation between \nife\ and \nafe\ has also been found for metal-poor
halo and thick-disk stars \citep{nissen10}.  As suggested by \citet{venn04},
the correlation can be explained if the  production of the most abundant
Ni isotope, $^{58}$Ni, and $^{23}$Na in Type II SNe has the same dependence on neutron
excess. If so, the scatter in the \nafe -age and \nife -age relations may be
related to variations in neutron excess of SNe belonging to the old disk population.

\subsection{3D non-LTE effects}
\label{3Dnon-LTE}
Because the stars in this paper were selected to lie on the main sequence and to
have effective temperatures in a small range (5600\,K $< \teff < 5950$\,K) we
expect only small differential 3D non-LTE corrections to the derived abundances
at a given \feh . There may, however, be significant
differential corrections when comparing metal-poor 
with metal-rich stars, which would affect the discontinuities
seen in  Fig. \ref{fig:xfe-age} at an age of $\sim 6$\,Gyr. 

As reviewed by \citet{barklem16}, non-LTE corrections based on modern physically-motivated 
descriptions of cross sections for collisions with hydrogen atoms and electrons 
are now available for several elements including Mg \citep{osorio16} and Al
\citep{nordlander17}. These corrections refer, however, to 1D hydrostatic model
atmospheres and don't take into account 3D effects on the spectral
lines. In 1D analyses, including that in the present paper, it is assumed that the
microturbulence broadening is independent of depth in the stellar atmosphere.
In real stellar atmospheres, the hydrodynamical broadening changes with depth and
to the extent that lines of different elements are formed at different depths, the derived
\xfe\ values may be in error \citep{ludwig16}. Obviously, the solution is 
to carry out a 3D model atmosphere analysis
in which line broadening due to atmospheric gas motions follows from the hydrodynamical
calculations \citep{asplund00, allende-prieto02}.

A non-LTE study of C and O lines based on ab initio calculations of cross sections
for collisions with hydrogen atoms tested via solar centre-to-limb observations
\citep{amarsi18, amarsi19a} have recently
been carried out by \citet{amarsi19b} for a grid of 3D model atmospheres.
The results are published as large tables providing 3D\,non-LTE\,$-$\,1D\,LTE corrections
as a function of 1D LTE values of \teff , \logg, \feh , \turb , and C or O abundance 
allowing interpolation to corrections for our set of stars. 
However, before presenting the results, a special problem with the method used 
to determine oxygen abundances needs to be addressed.

\begin{figure}
\resizebox{\hsize}{!}{\includegraphics{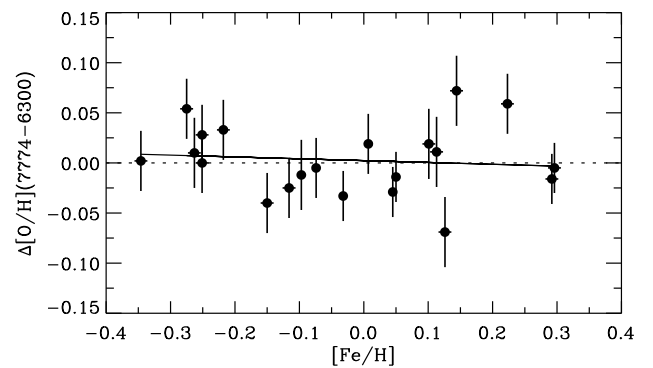}}
\caption{Difference between oxygen abundances derived from the 7774\,\AA\ \OI\
triplet and the \oI\ 6300\,\AA\ line versus \feh\ after applying 3D non-LTE
corrections from \citet{amarsi19b}. The full drawn line is a least-squares
fit to the data.}
\label{fig:O.7774-6300}
\end{figure}

The \OI\ triplet at 7774\,\AA\ is not covered by the HARPS spectra,
so we used the weak \oI\ line at 6300.3\,\AA\ to determine O abundances.
As described in Paper\,I, correction for a blending \NiI\ line 
\citep{allende-prieto01} was applied
by calculating its EW in the solar and stellar spectra 
using Ni abundances derived from other \NiI\ lines and an oscillator strength
(log\,$gf = -2.11$) measured by \citet{johansson03}. This procedure could 
cause systematic errors in the derived O abundances of the more metal-rich stars. 
The Ni line makes up a larger and larger fraction of the total EW of the 6300.3\,\AA\ 
blend as \feh\ is increasing, because \nife\ is increasing with \feh\ \citep{adibekyan12, bensby14}
while \ofe\ is decreasing \citep{amarsi19b}. In order to test this problem we have 
determined oxygen abundances from the \OI\ 7774\,\AA\ triplet lines for a subset
of stars for which reasonable precise EWs can be measured from 
ESO/FEROS spectra having $R = 48\,000$ and $S/N \sim 200$.
For 11 of our stars, EWs of the \OI\ triplet lines
can be found in \citet{nissen14} and for another 10 stars
EWs were measured from FEROS spectra that have become available in the ESO Science Archive 
after 2014. These EWs were used to derive 1D LTE values of \oh\,
and after applying 3D non-LTE corrections from \citet{amarsi19b}
a comparison was made with 3D non-LTE values of \oh\ based on the \oI\ line 
at 6300.3\,\AA\ (see Table \ref{table:Oabun}). As seen from Fig. \ref{fig:O.7774-6300}, 
there is a satisfactory agreement between the
two sets of \oh\ values. In particular, there is no trend of the difference 
as a function of \feh\ indicating that the oxygen abundances derived from the 
forbidden line can be trusted. In this connection, we note that whereas the
differential 3D non-LTE corrections are almost negligible for the \oI\ line,
they range from $-0.06$\,dex to +0.05\,dex for the triplet.

\begin{table}
\caption[ ]{Oxygen abundances derived from the \OI\ 7774\,\AA\ triplet and from
the \oI\ line at 6300\,\AA .}
\label{table:Oabun}
\centering
\setlength{\tabcolsep}{0.10cm}
\begin{tabular}{rrrrrr}
\noalign{\smallskip}
\hline\hline
\noalign{\smallskip}
\noalign{\smallskip}
  HD no.&  \ohyy   & \ohyy   & \ohxx   & \ohxx   & $\Delta \oh$\,\tablefootmark{a} \\
        &   1D,LTE &  3D,nLTE & 1D,LTE  & 3D,nLTE &   3D,nLTE    \\ 
\noalign{\smallskip}
\hline
\noalign{\smallskip}
   11505&   0.037 &  0.020  &  0.029  & 0.020  &  0.000 \\
   19467&   0.072 &  0.048  &  0.096  & 0.088  & $-$0.040 \\
   20766&  $-$0.127 & $-$0.085  & $-$0.115  &$-$0.118  &  0.033 \\
   20782&  $-$0.048 & $-$0.061  & $-$0.054  &$-$0.056  & $-$0.005 \\
   45184&  $-$0.002 & $-$0.017  & $-$0.004  &$-$0.003  & $-$0.014 \\
   45289&   0.101 &  0.071  &  0.110  & 0.104  & $-$0.033 \\
   89454&   0.098 &  0.115  &  0.043  & 0.043  &  0.072 \\
   96423&   0.055 &  0.046  &  0.038  & 0.035  &  0.011 \\
  102365&  $-$0.194 & $-$0.147  & $-$0.143  &$-$0.149  &  0.002 \\
  108309&   0.128 &  0.066  &  0.141  & 0.135  & $-$0.069 \\
  114853&  $-$0.211 & $-$0.175  & $-$0.182  &$-$0.185  &  0.010 \\
  117207&   0.158 &  0.153  &  0.099  & 0.094  &  0.059 \\
  126525&  $-$0.100 & $-$0.068  & $-$0.054  &$-$0.056  & $-$0.012 \\
  134664&   0.010 & $-$0.002  &  0.017  & 0.017  & $-$0.019 \\
  134987&   0.232 &  0.198  &  0.221  & 0.214  & $-$0.016 \\
  146233&  $-$0.008 & $-$0.011  &  0.018  & 0.018  & $-$0.029 \\
  160691&   0.239 &  0.188  &  0.200  & 0.193  & $-$0.005 \\
  189567&  $-$0.092 & $-$0.073  & $-$0.122  &$-$0.127  &  0.054 \\
  202628&  $-$0.080 & $-$0.069  & $-$0.091  &$-$0.088  &  0.019 \\
  210918&   0.006 & $-$0.009  &  0.021  & 0.016  & $-$0.025 \\
  211415&  $-$0.172 & $-$0.170  & $-$0.194  &$-$0.198  &  0.028 \\
\noalign{\smallskip}
\hline
\end{tabular}
\tablefoot{
\tablefoottext{a}{The 3D non-LTE value of the difference between \ohyy\ and \ohxx .}}
\end{table}

The differential 3D non-LTE corrections relative to those for the Sun 
range from $-0.015$\,dex to +0.005\,dex for the  
carbon abundances derived from high-excitation \CI\ lines 
and from $-0.011$\,dex to +0.003\,dex for the oxygen abundances derived 
from the \oI\ line at 6300.3\,\AA . These small corrections do not lead to any
significant change of the 1D LTE trends of \cfe\ and \ofe\ with age
shown in Fig. \ref{fig:xfe-age}. The 3D non-LTE effects on the iron
abundances may, however, be more important. According to \citet{amarsi19b},
the \feh\ values derived from our \FeII\ lines are subject to changes 
ranging from $-0.007$\,dex to +0.025\,dex,
but 3D non-LTE corrections for the \FeI\ lines used to determine the atmospheric
parameters are not available. Obviously, a 3D non-LTE study of the formation 
of both \FeI\ and \FeII\ lines in the atmospheres of solar-type stars as already
carried out for the Sun \citep{lind17} is needed before improved \feh\ values can be obtained.

\begin{figure}
\resizebox{\hsize}{!}{\includegraphics{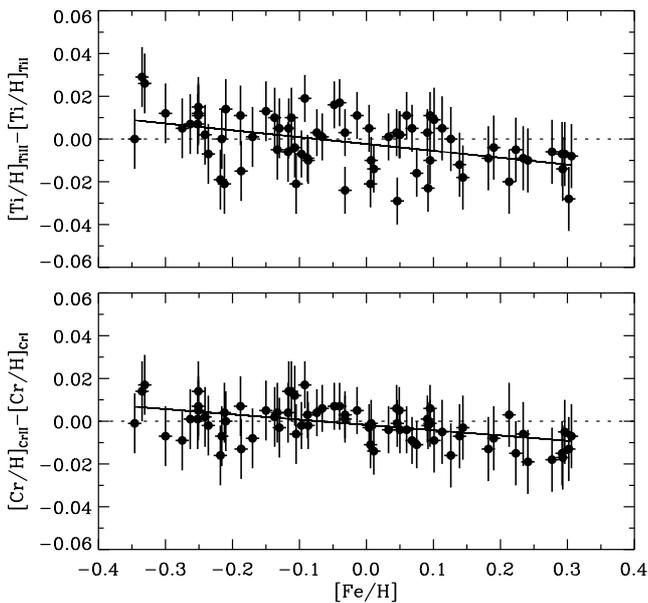}}
\caption{Comparison of Ti and Cr abundances derived from  lines belonging to
the neutral and ionised species.}
\label{fig:TiIICrII-feh}
\end{figure}

In the case of Ti and Cr, we have determined abundances from respectively
three and two lines of the majority
species \TiII\ and \CrII\ in addition to the abundances given in
Table \ref{table:abun} that are based on respectively nine \TiI\
and seven \CrI\ lines. As seen from Fig. \ref{fig:TiIICrII-feh}, the abundances
from the ionised lines tend to deviate from those of the neutral lines
by $\sim +0.01$\,dex at the lowest metallicities and $\sim -0.01$ at high \feh .
This suggests that the 1D LTE results shown in Fig. \ref{fig:xfe-age} 
may be affected by differential 3D non-LTE effects at a level of $\pm 0.01$  
when comparing low- and high-metallicity stars. Again, we need 3D non-LTE
studies to improve the accuracy of abundances of these and other elements.

\subsection{C/O versus stellar age}
\label{COratio}
The C/O number ratio is important for the structure and composition of planets.
The Sun has  $\conum = 0.56 \pm 0.05$ \citep{amarsi19a}, but if
a star and its proto-planetary disk have $\conum \simgt 0.8$, `Carbon' planets
containing large amounts of graphite and carbides instead of Earth-like silicates
may be formed \citep{kuchner05, bond10}. Work by \citet{delgado-mena10} and
\citet{petigura11} suggested that a significant percentage of 
high-metallicity solar-type stars
have $\conum > 0.8$, but later studies have not confirmed this
\citep{nissen13, brewer16, suarez-andres18, amarsi19b, stonkute20}. According to
these works, there is a rise of C/O with \feh , but none of the
stars have C/O significantly larger than 0.8. 

\begin{figure}
\resizebox{\hsize}{!}{\includegraphics{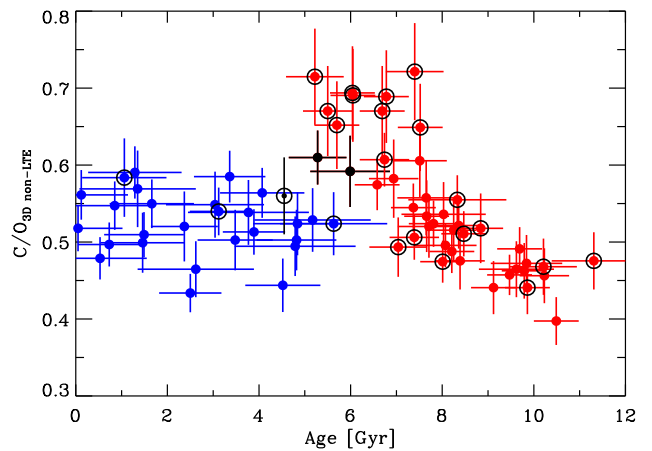}}
\caption{The C/O number ratio as a function of stellar age
with blue and red filled circles referring to the two
sequences Fig. \ref{fig:feh-age}. Stars for which one or more
planets have been detected are labeled with a black ring.}
\label{fig:CO.ratio-age}
\end{figure}

As seen from Fig. \ref{fig:CO.ratio-age}, our 3D non-LTE values of C/O
show a rising trend with decreasing age for the old (red) sequence, but
all stars have C/O clearly below 0.8. Interestingly, all nine stars
with $\conum \simeq 0.7$ have confirmed planets according to The Extrasolar Planets
Encyclopaedia\,\footnote{{\tt http://exoplanet.eu}} \citep{schneider12}.
These stars have also high metallicities, i.e. $0.18 < \feh < 0.31$,
so it cannot be decided from the present small sample whether it is
the high C/O ratio or the high \feh\ (or both) that favour
formation of planets.

\subsection{\ymg\ and \srmg\ versus age}
\label{YMg-age}
As mentioned in the Introduction several studies of solar twins
suggest that \ymg\ can be used as a sensitive chemical clock to obtain stellar ages
\citep{nissen16, tucci-maia16, spina18}. \citet{feltzing17} found, however,
that \ymg\ at a given age decreases significantly with decreasing metallicity,
but this is less obvious from data presented by \citet{delgado-mena19} and \citet{titarenko19}.

\begin{figure}
\resizebox{\hsize}{!}{\includegraphics{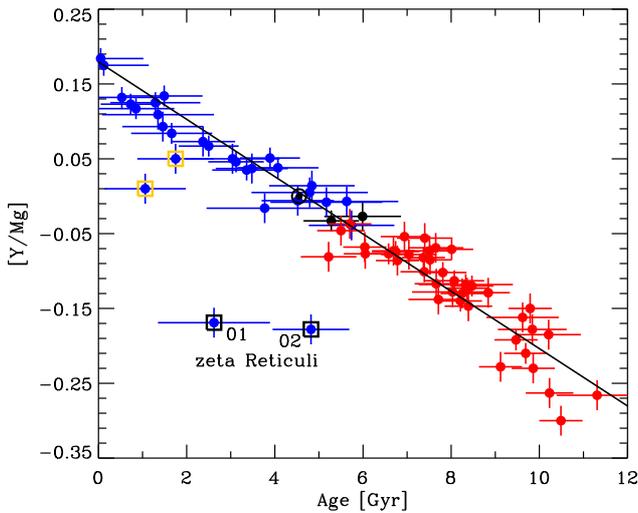}}
\caption{\ymg\ versus stellar age with the same symbols
as in Fig. \ref{fig:feh-age}. The line corresponds to the linear fit given in Eq. (6).}
\label{fig:ymg-age}
\end{figure}

Figure \ref{fig:ymg-age} shows a tight \ymg -age relation for our
sample of stars except for a strong deviation of the two components
of the visual binary star, $\zeta$\,Reticuli, and a minor deviation of the
two Na rich stars. Excluding these four stars, an error-weighted maximum likelihood fit
gives
\begin{eqnarray}
\ymg = 0.179 \, (\pm 0.007) - 0.0383\, (\pm 0.0010) \cdot {\rm Age} \,{\rm [Gyr]}, 
\end{eqnarray}
with a reduced chi-square of  0.82. Within the quoted errors of the zero-point 
and the slope, this age calibration of \ymg\ agrees with the calibration 
in Paper\,II for 21 solar twins, whereas \citet{spina18} find a somewhat different slope,
i.e. $-0.045 \pm 0.002$ for 76 solar twins, which may be due to a different 
age scale, i.e. the use of Yonsei-Yale isochrones \citep{yi01, kim02} instead
of the ASTEC models used by us (see discussion in Paper\,II).

\begin{figure}
\resizebox{\hsize}{!}{\includegraphics{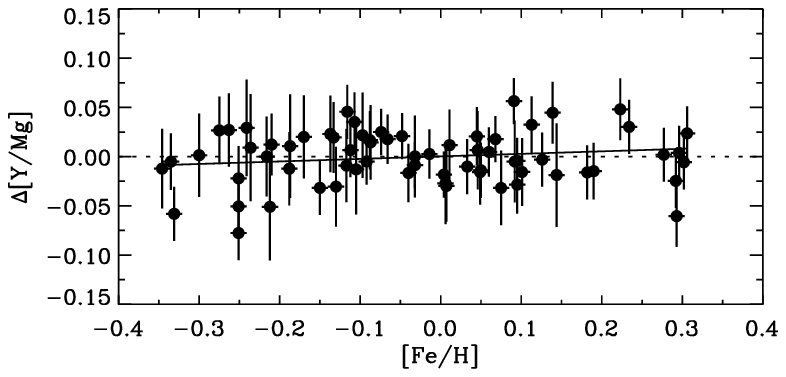}}
\caption{Deviations of \ymg\ from Eq. (6) as a function of \feh .
The line shows a linear fit to the data.}
\label{fig:deltaymg-feh}
\end{figure}

As seen from Fig. \ref{fig:ymg-age} there is no indication of a shift in
the \ymg -age relation between the old (red) and the younger (blue) sequence of stars.
Furthermore, Fig. \ref{fig:deltaymg-feh} shows that the residuals of \ymg\
from the age calibration given by  Eq. (6) has no significant dependence on metallicity
in the $-0.3 < \feh < +0.3$ range. In this connection, we note that the shift of \ymg\
with \feh\ found by \citet{feltzing17} occurs for stars with $\feh < -0.3$, so 
there is not necessarily a contradiction between the results  of the two investigations.

\begin{figure}
\resizebox{\hsize}{!}{\includegraphics{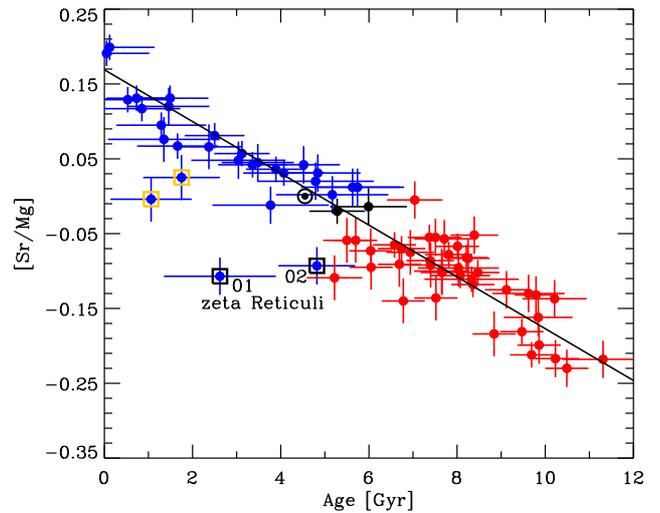}}
\caption{\srmg\ versus stellar age with the same symbols
as in Fig. \ref{fig:feh-age}. The line shows the linear fit given in Eq. (7).}
\label{fig:srmg-age}
\end{figure}

\begin{figure}
\resizebox{\hsize}{!}{\includegraphics{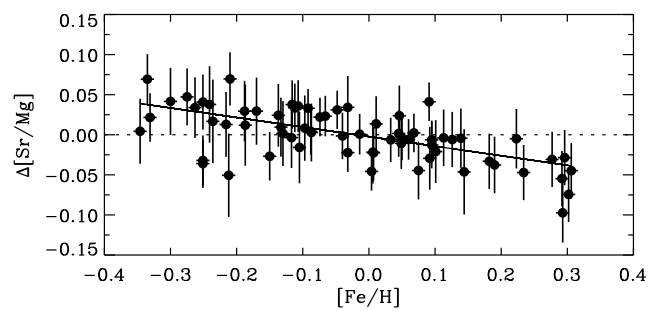}}
\caption{Deviations of \srmg\ from Eq. (7) as a function of \feh .
The line shows the linear fit to the data.}
\label{fig:deltasrmg-feh}
\end{figure}

Strontium ($Z = 38$) belongs to the first peak of $s$-process elements
like yttrium ($Z = 39$). According to \citet{karakas16}, the yields of
these two elements from Asymptotic Giant Branch (AGB) stars depend
in a similar way on stellar mass and metallicity, so one would expect 
\srmg\ to have about the same dependence on age as \ymg . As shown in
Fig. \ref{fig:srmg-age}, this is indeed the case; the error-weighted fit
to the data gives
\begin{eqnarray}
\srmg = 0.169 \, (\pm 0.007) - 0.0346\, (\pm 0.0010) \cdot {\rm Age} \,{\rm [Gyr]}, 
\end{eqnarray}
when excluding $\zeta$\,Reticuli and the two Na-rich stars. The scatter around
this line is, however, larger than in the case of \ymg\ and  there is
a significant dependence of the residuals on \feh\  
as seen from Fig. \ref{fig:deltasrmg-feh}.
In this connection, we note that the Sr abundances were determined from only one 
\SrI\ line ($\chi_{\rm exc.} = 0.0$\,eV) at 4607.3\,\AA , 
which according to \citet{grevesse15} is
subject to a solar non-LTE correction of +0.15\,dex based on calculations by 
M. Bergemann (priv. communication). This suggests that there may be
significant differential 3D non-LTE corrections depending on \feh .
Yttrium abundances, on the other hand, were determined from three lines.
All belong to the \YII\ majority species and are thus not susceptible 
to non-LTE overionisation effects.
We conclude that until 3D non-LTE calculations
become available, \ymg\ is to be preferred as a chemical clock.

As recently discussed by \citet{jofre20}, chemical abundances for 80 solar twins
\citep{spina18, bedell18}, suggest that a variety of abundance ratios between
$s$-process elements (from Sr to Ce) and lighter elements (from C to Zn) 
can be used as sensitive chemical clocks. It remains, however, to be investigated  
to which extent these ratios depend on \feh\ at a given age. As an example,
our data show that \yal\ is more sensitive to age than \ymg ,
but the residuals in the fit depend significantly on \feh\ as in
the case of \srmg . 

\subsection{The $\zeta$ Reticuli binary star}
\label{zetaRet}
The binary star $\zeta$\,Ret in the constellation Reticulum 
is in many ways a most remarkable system. The two components,
$\zeta ^1$\,Ret (HD\,20766) and $\zeta ^2$\,Ret (HD\,20807) have apparent magnitudes of 
$V = 5.51$ and $V = 5.23$, respectively, and with a separation of 5.15 arc minutes
they can been seen with the naked eye on a dark sky. 
According to $Gaia$ DR2 data \citep{gaia.collaboration18}, their
proper motions, radial velocities, and distances (12.0\,pc) agree, 
suggesting that they have a common origin.

\begin{figure}
\resizebox{\hsize}{!}{\includegraphics{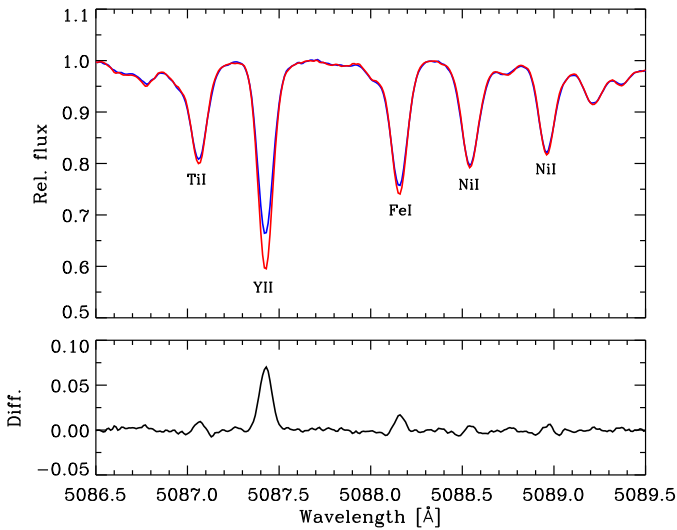}}
\caption{HARPS spectrum of $\zeta ^1$\,Ret (blue line) around the \YII\ line at
5087.4\,\AA\ in comparison with the HARPS spectrum of HD\,20619 (red line). 
The lower panel shows the difference $\zeta ^1$\,Ret -- HD\,20619.}
\label{fig:5087}
\end{figure}

As seen from Fig. \ref{fig:ymg-age}, both $\zeta ^1$ and $\zeta ^2$\,Ret fall
much below the \ymg -age relation for the other stars. This is due to an unusually low
yttrium abundance compared to other stars with similar ages as shown in Fig. \ref{fig:5087}, 
where the spectrum of $\zeta ^1$\,Ret is compared to that of HD\,20619. The 
two stars have about the same values of \teff , \logg , and \feh\
and the strengths of the \TiI , \FeI , and \NiI\ lines are nearly the same,
whereas the \YII\ line at 5087.4\,\AA\ is weaker in the spectrum 
of $\zeta ^1$\,Ret resulting in a deficiency, $\Delta \yfe = -0.13$\,dex, 
of $\zeta ^1$\,Ret relative to HD\,20619.
 
\begin{figure}
\resizebox{\hsize}{!}{\includegraphics{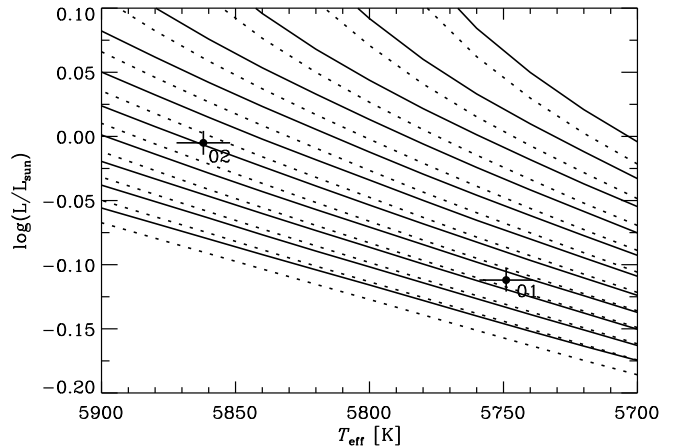}}
\caption{$\zeta ^1$ and $\zeta ^2$\,Ret  in the
log\,($L / L_{\rm sun}$) versus \teff\ diagram in comparison with ASTEC isochrones 
ranging in age from 0 to 10\,Gyr in steps of 1\,Gyr. 
The full drawn lines refer to isochrones interpolated to
the surface heavy element abundance of $\zeta ^1$, $Z_s = 0.0122$, and the dotted lines
to isochrones corresponding to the abundance of $\zeta ^2$\,Ret, $Z_s = 0.0111$.} 
\label{fig:iso.zetaRet}
\end{figure}

In order to illustrate how precisely the ages of $\zeta ^{1,2}$\,Ret are  
determined, Fig. \ref{fig:iso.zetaRet} shows their locations in the 
log\,($L / L_{\sun}$)-\,\teff\ diagram in comparison with ASTEC
isochrones ranging in age from 0 to 10\,Gyr. As the metallicities 
and \alphafe\ values of the two stars are slightly different, 
two sets of isochrones corresponding to the heavy element abundances 
of the stars are shown.
Taking into account the estimated uncertainties of \teff\ and 
log\,($L / L_{\sun}$) shown with error bars in the figure as well as the
uncertainty of the $Z_s$-values of the stars we obtain an age of
$2.6 \pm 1.3$\,Gyr for $\zeta ^1$\,Ret and $4.7 \pm 0.9$\,Gyr for $\zeta ^2$\,Ret. 
Thus, there is a hint of a difference, $\Delta {\rm Age} = 2.1 \pm 1.6$\,Gyr,
between the ages but this is only significant at the 1.3-sigma level.

\begin{figure}
\resizebox{\hsize}{!}{\includegraphics{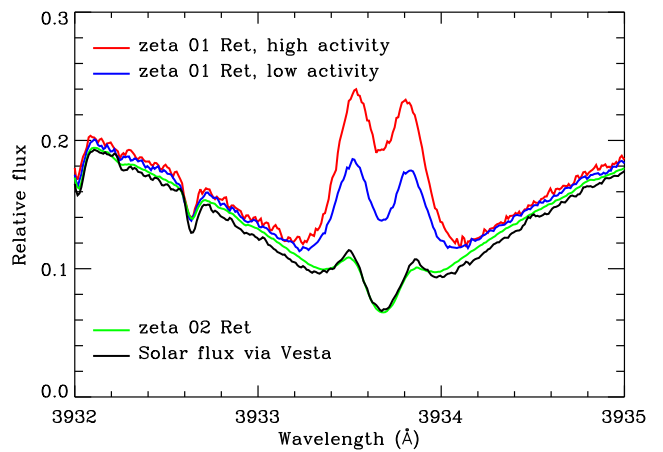}}
\caption{HARPS spectra of $\zeta ^1$ and $\zeta ^2$\,Ret near the
centre of the \CaII\ K line in comparison with the solar flux HARPS spectrum 
observed as reflected light from Vesta in 2011.
The flux is given relative to the continuum flux in the region
of the \CaII\ H\&K lines.}
\label{fig:CaIIK}
\end{figure}

Ages of $\zeta ^{1,2}$\,Ret may also be estimated from their 
chromospheric activity. From a detailed study of the \CaII\ H\&K lines
in HARPS spectra, \citet{flores18} found a period of $\sim \! 10$\,yr in the
activity of  $\zeta ^2$\,Ret with approximately the same level of \CaII\ H\&K core
emission  as the Sun. $\zeta ^1$\,Ret is much more
active as can be seen from Fig. \ref{fig:CaIIK}. No  
activity cycle was  detected by \citet{flores18} but the emission 
in the cores of the \CaII\ H\&K lines varies
over the 3.3\,yr period covered by the HARPS spectra. 
The average chromospheric \CaII\ H\&K emission index 
($R' _{\rm HK} = L_{\rm HK} / L_{\rm bol}$) determined for the 
two stars \citep[see Table 2 in][]{flores18}
is log\,$R' _{\rm HK} = -4.68 \pm 0.04$ for $\zeta ^1$\,Ret
and log\,$R' _{\rm HK} = -4.86 \pm 0.02$ for $\zeta ^2$\,Ret.
From these values and the open cluster age calibration of log\,$R' _{\rm HK}$
by \citet{mamajek08}, which has a scatter of $\pm 0.07$\,dex in log\,$R' _{\rm HK}$,
we estimate ages of $1.8 \pm 0.8$\,Gyr for $\zeta ^1$\,Ret and 
$4.10 \pm 1.1$ for $\zeta ^2$\,Ret in agreement with the isochrone ages
within the estimated errors. 

The chemical evolution ages of the $\zeta$\,Ret stars are significantly
higher than their isochrone and activity ages.
From the \ymg\ values we derive 
$9.1 \pm 0.5$\,Gyr for $\zeta ^1$\,Ret and $9.4 \pm 0.5$\,Gyr for $\zeta ^2$\,Ret
based on the calibration in Eq. (6). A similar problem
was presented by \citet{rocha-pinto02}, who noted that $\zeta ^1$\,Ret is
chromospherically young but kinematically old. Its Galactic velocity
components with respect to the Local Standard of Rest (LSR) are 
($U, V, W$) = ($-63, -34, +23$)\,\kmprs , indicating that it belongs to the old disk.
As an explanation, \citet{rocha-pinto02} suggested that such stars are 
blue stragglers formed from the coalescence of short-period binaries with 
low-mass components. This scenario also explains the difference between 
isochrone age and chemical evolution age of $\zeta ^{1,2}$\,Ret. It may even explain
the possible difference in isochrone age between $\zeta ^1$ and $\zeta ^2$\,Ret, 
because the coalescence could happen at different times. Furthermore, the scenario
can explain that both stars are severely depleted in lithium and
beryllium \citep{santos04} relative to stars with similar mass, age, and metallicity,
because Li and Be destruction would take place in the deep convection zones 
of the low-mass stars.

\begin{figure}
\resizebox{\hsize}{!}{\includegraphics{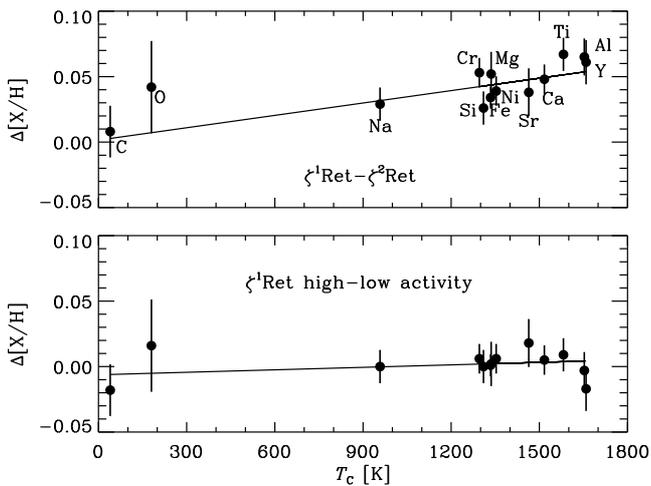}}
\caption{Abundance difference between $\zeta ^1$
and $\zeta ^2$\,Ret as a function of elemental condensation temperature
(upper panel) compared to the difference in abundances derived for $\zeta ^1$\,Ret
in respectively the high and low activity phases shown in Fig. \ref{fig:CaIIK} (lower panel).}
\label{fig:deltaxh-Tc.zetaRet}
\end{figure}

It is more difficult to explain that  $\zeta ^1$ and $\zeta ^2$\,Ret do not
have the same chemical composition. \citet{saffe16} used HARPS
spectra with $S/N \sim 300$ obtained on a single night to 
show that abundances of refractory elements in $\zeta ^1$\,Ret are 
enhanced relative to the abundances in $\zeta ^2$\,Ret. Furthermore, they 
found that the abundance difference increases as a function of elemental condensation
temperature (\Tc ) \citep{Lodders03} with a slope of $3.85 \, (\pm 1.02) \cdot  10^{-5}$dex\,K$^{-1}$.
This was confirmed by \citet{adibekyan16} based on combined HARPS spectra
of $\zeta ^1$ and $\zeta ^2$\,Ret having S/N ratios as high as 1300 and 3000,
respectively. As seen from Fig. \ref{fig:deltaxh-Tc.zetaRet}, our abundances
also show a difference between  $\zeta ^1$ and $\zeta ^2$\,Ret correlated
with elemental condensation temperature;
a weighted least squares fit leads to the relation
\begin{eqnarray}
\Delta \xh = 0.002 (\pm 0.015) + 3.15 \, (\pm 1.13) 10^{-5} \cdot  \Tc \, {\rm dex\,K}^{-1},
\end{eqnarray}
in agreement with the results of \citet{saffe16} and \citet{adibekyan16}.

\citet{galarza19} have shown that some of the stronger  
\FeI\ and \FeII\ lines in HARPS spectra of the young active
solar twin HD\,59967 vary significantly in strength along its
activity cycle of $\sim \! 6$\,yr and that the derived atmospheric parameters 
change correspondingly with an amplitude of $\sim \! 30$\,K in
\teff\ and 0.015\,dex in \feh . A similar result was obtained by \citet{spina20}
for a large sample of active solar-type stars.
This raises the question if our derived
abundances for $\zeta ^1$\,Ret also depend on its level of activity.
We have therefore analysed  two sets of HARPS spectra, each with $S/N \sim 800$,
obtained during respectively the high and low activity phases 
(see Fig. \ref{fig:CaIIK}).
The derived \teff 's differ by 10\,K only and as shown in 
Fig. \ref{fig:deltaxh-Tc.zetaRet} (lower panel),
there is no significant difference between the abundances obtained
during the two activity phases. This supports that our derived abundances are 
not significantly affected by magnetic activity and that the abundance
enhancement of $\zeta ^1$\,Ret relative to $\zeta ^2$\,Ret cannot
be explained by the difference in activity. We note in this connection that
$\zeta ^1$\,Ret is not as active as HD\,59967 and that we are using
weaker Fe lines less sensitive to magnetic fields than those used by 
\citet{galarza19} and \citet{spina20} to determine atmospheric parameters and abundances.

As recently listed by \citet[][Table 6]{ramirez19}, there is now ten known twin-star
comoving pairs with a significant difference 
in the abundance of refractory elements. In some cases the
difference reaches $\sim \! 0.2$\,dex \citep{oh18, ramirez19, nagar20}
and is correlated with elemental condensation temperature.
Possible explanations  
of the abundance differences include sequestration of refractory elements
in planets \citep{melendez09, chambers10}, engulfment of planets
\citep[e.g.][]{melendez17, tucci-maia19}, and dust-gas separation
in star-forming gas clouds or circumstellar disks \citep{gaidos15,
gustafsson18a, gustafsson18b}. In this connection we note that no planets
have been detected around the $\zeta$\,Ret stars, but infrared
observations with the {\em Spitzer} and {\em Herschel} telescopes suggested the
existence of a debris disk around $\zeta ^2$\,Ret
\citep{Trilling08, Eiroa10}. ALMA/Atacama Compact Array (sub)mm 
observations carried out 8 years after the $Herschel$ observations show,
however, that the disk does not share the large proper motion of $\zeta ^2$\,Ret
and is probably a background source \citep{faramaz18}. Therefore, it remains
unclear how the abundance difference between $\zeta ^1$ and $\zeta ^2$\,Ret
should be explained.

\section{Discussion}
\label{discussion}
When discussing explanations of the two sequences for \feh\
and \xfe\ as a function of stellar age, it should be remembered that
the stars have been selected to have an approximately uniform
distribution in \feh\ from $\sim -0.3$\,dex to $\sim +0.3$\,dex.  
For this range the separation in low-
and high-alpha stars is not as clear as in the range
$-0.6 \simlt \feh \simlt -0.3$ \citep[e.g.][]{adibekyan12, bensby14, fuhrmann17}.
Instead, we have found indication of a dichotomy of
the distribution of stars in the \feh -age diagram. 

One may wonder why the two age sequences have not been seen in previous studies 
of the age-metallicity relation for solar-type stars, but for typical
age precisions of 2-4\,Gyr, the dearth of stars at intermediate ages
seen in Fig. \ref{fig:feh-age} tends to be washed out. A higher age precision was
claimed by \citet{haywood13}, who in a study of the
\citet{adibekyan12} sample  estimated errors of 0.8-1.5\,Gyr for
isochrone ages of somewhat evolved main-sequence stars. Interestingly, they 
found a tight age-metallicity relation for high-alpha
(thick disk) stars similar to the old sequence in our Fig. \ref{fig:feh-age},
but there is no split into two distinct age sequences in their age-metallicity
diagram \citep[see Fig. 9 in][]{haywood13}. The reason may be that Haywood et al.
include stars from a broad range in effective temperature, 
5000\,K $\simlt \teff \simlt 6400$\,K, which means that systematic errors in the
isochrones as a function of stellar mass add to the statistical errors.
Our study is limited to stars with 5600\,K $< \teff < 5950$\,K allowing us
to make a differential determination of isochrone ages relative to the 
well known age of the Sun and therefore obtain ages with a precision of 0.5-1.3\,Gyr
for stars with similar \feh . Furthermore, the precision of our derived
effective temperatures is on the order of 10\,K compared to 25-70\,K for
the stars in \citet{adibekyan12}, which is important for the derivation 
of precise ages especially for stars near the ZAMS. 

\begin{figure}
\resizebox{\hsize}{!}{\includegraphics{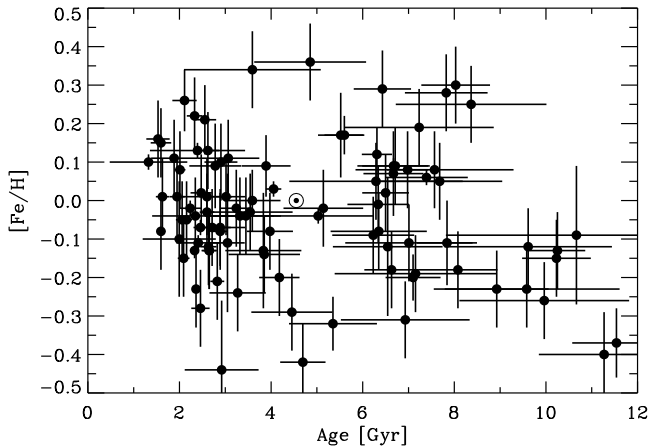}}
\caption{Age-metallicity diagram for dwarf stars with
ages determined from {\em Kepler} short-cadence observations 
of individual oscillation frequencies by \citet{silva-aguirre15, silva-aguirre17}.}
\label{fig:victor}
\end{figure}

Concerning investigations based on asteroseismic ages,
we note that the analysis by \citet{silva-aguirre18} of $\sim \! 1200$ 
red giants with precise abundances from APOGEE (the so-called
APOKASC sample)  does not show any split of 
the age-metallicity relation into two sequences. 
The seismic ages have, however, errors of about 20-40\,\%
depending on whether a star belongs to the red giant branch or the clump phase.
This means that stars with ages in the range 5-10\,Gyr have typical age uncertainties 
of 1.5-3\,Gyr, so again one cannot expect to detect the dearth of stars with 
intermediate ages seen in Fig. \ref{fig:feh-age}. The smaller sample of solar-type dwarfs  
for which high-precision individual oscillation frequencies are available
from {\em Kepler} short-cadence observations \citep{huber13, lund17}
is more interesting in this connection. \citet{silva-aguirre15} analysed 33
such stars with planets and \citet{silva-aguirre17} 66 stars
without known planets, the so-called LEGACY sample. They obtained 
age uncertainties of 10-15\,\% from the asteroseismic data. Metallicities
are, on the other hand, not very precise for many of these stars; 
ten stars have \feh\ determined from high S/N HARPS-N spectra 
with a precision of $\sim \! 0.02$\,dex \citep{nissen17}, but for the rest 
\feh\ was derived from low S/N spectra with
a precision of $\sim \! 0.10$\,dex. Nevertheless, the age-metallicity 
diagram for the {\em Kepler} dwarfs (see Fig. \ref{fig:victor}) gives support
to the existence of an old and a young sequence with a deficit of stars 
in between, although details such as the \feh - age slope for the two 
populations and the width of their separation are somewhat different from
the corresponding features in the ASTEC age-metallicity diagram. The deficit of stars 
at intermediate ages was in fact noted 
by \citet{silva-aguirre15} but was ascribed to a selection effect, namely that 
detection of oscillations by {\em Kepler} is favoured for the brighter stars,
either young F-type or old evolved G-type stars. After
addition of the LEGACY sample, Fig. \ref{fig:victor} includes, however, 
ten unevolved ($4.27 < \logg < 4.50$), cool (5500\,K $< \teff < 6150$\,K
stars with ages corresponding to those of the young sequence. 
Stars with intermediate ages are more evolved and have larger oscillation
amplitudes. Therefore, it is unlikely that the deficit of stars at intermediate ages 
is a selection effect.  
The absence of stars younger than $\sim \! 1.5$\,Gyr is on the other hand a selection bias;
Such stars lie either close to the ZAMS and have low oscillation amplitudes or
are warmer than $\teff \sim 6600$\,K, the temperature limit of the {\em Kepler}
dwarf sample.

As a possible explanation of the two age sequences in  Fig. \ref{fig:feh-age}
we have considered the ''two-infall" chemical evolution model of 
\citet{chiappini97} as revised by \citet{grisoni17}. This model assumes 
two main episodes of infalling gas onto
the Galactic disk corresponding to the formation of respectively the
thick and the thin disk. \citet{spitoni19, spitoni20} have recently compared the
predictions of this model with APOGEE chemical abundances and seismic ages of 
K giants in the APOKASC sample \citep{silva-aguirre18} and have
shown that a significant delay of 4.5 to 5.5\,Gyr between the two episodes of 
gas accretion is needed to explain the data. The corresponding star formation
rate has a minimum at an age of $\sim \! 8$\,Gyr. A similar quenching of
star formation around an age of 8\,Gyr was derived by \citet{snaith15} from 
the \citet{adibekyan12} abundances and the \citet{haywood13} isochrone ages of
solar-type stars. There is also indications of a double-peaked star formation 
history, although with a minimum around 6\,Gyr,
from $Gaia$ colour-magnitude diagrams \citep{mor19}. Furthermore, several
models of galaxy formation predict a quenching of star formation 
at ages between 6 and 9\,Gyr. \citet{noguchi18} suggests that
high-alpha stars form during an initial phase of accretion of cold
primordial gas followed by a hiatus until the shock-heated gas has cooled 
due to radiation and a new accretion begins leading to the formation of
low-alpha disk stars. Cosmological hydrodynamical simulations  
also point to the formation of bimodal disks \citep[e.g.][]{grand18} 
possibly triggered by a gas-rich merging satellite \citep{buck20}.

\begin{figure}
\resizebox{\hsize}{!}{\includegraphics{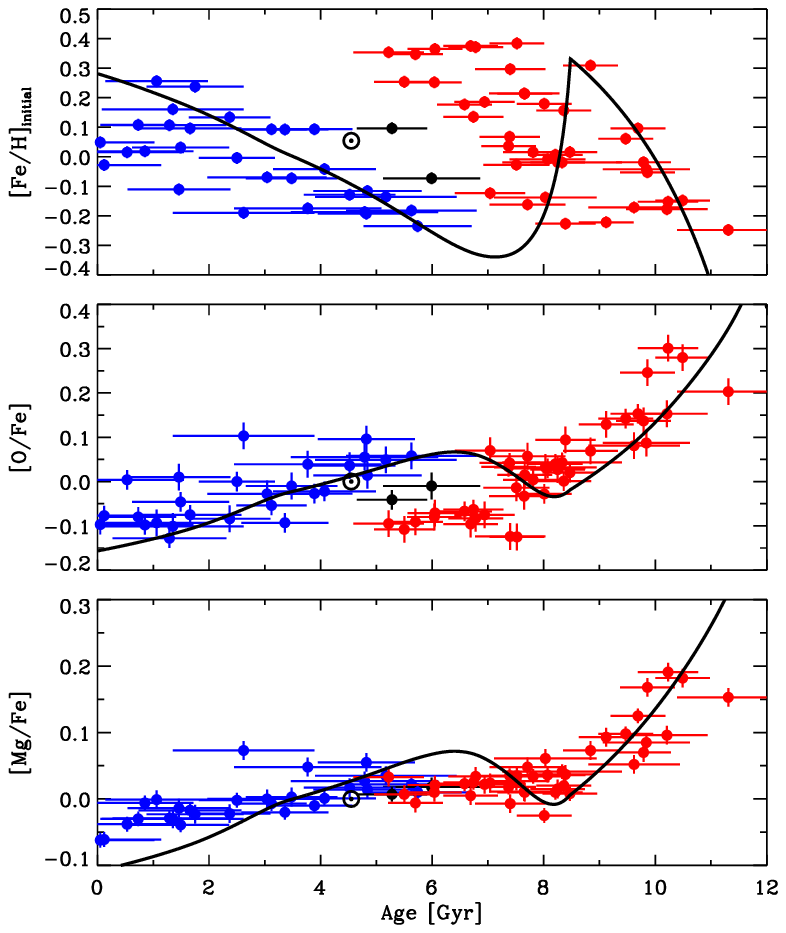}}
\caption{Comparison of the \fehi -, \ofe -, and \mgfe -age relations with
predictions from the two-infall chemical evolution model.}
\label{fig:spitoni}
\end{figure}

Using the Markov Chain Monte Carlo (MCMC) method described by 
\citet{spitoni20} we have made a preliminary investigation of how well the   
two-infall model fits the observed \feh -, \ofe -, and \mgfe -age relations.
The solar neighbourhood is assumed to have a metallicity of
$\feh = -1$ at an age of 12\,Gyr at which time the first exponentially decaying
infall of primordial gas begins followed by a delayed second infall of
primordial gas. The main parameters of the model are the timescales of the
infalling gas, $\tau_1$ and $\tau_2$, and the delay time, $T_{\rm delay}$.
More details about the model may be found in \citet{spitoni19, spitoni20}
including assumptions about the star formation efficiency, stellar yields, 
the Initial Mass Function (IMF), and the delay of Type\,Ia SNe relative to Type\,II SNe.
 
The two-infall model describes the evolution of chemical abundances in
interstellar gas and should therefore be compared
to the initial composition of stars. Hence, the present observed surface
abundances should be corrected for diffusion before comparison 
with the model predictions. According to the ASTEC models, the difference \fehi $-$\feh\ 
is 0.054\,dex for the Sun, rises to about 0.10\,dex for $\sim \! 10$\,Gyr old stars,
and is near zero for the youngest stars in our sample. Therefore, the age 
slope of \fehi\ is somewhat different from the slope of \feh . This
has been taken into account in the MCMC fitting of the two-infall model
to our data. The diffusion effect on \ofe\ and \mgfe\ may, on the other hand,
be neglected, because the abundances of these elements are affected in approximately the
same way \citep[see][Fig. 14]{turcotte98}.

Figure \ref{fig:spitoni} shows the MCMC fitting of the two-infall
model to our data. The resulting parameters values are $\tau_1 = 0.38 \pm 0.04$\,Gyr,
$\tau_2 = 3.2 \pm 0.4$\,Gyr, and $T_{\rm delay} = 3.52 \pm 0.05$\,Gyr. 
The delay time is similar to the value obtained from the MCMC fitting  
of the APOKASC data by \citet{spitoni20} ($T_{\rm delay} = 4.62 \pm 0.12$\,Gyr),
but the infall timescales are shorter than the values obtained for 
the APOKASC sample ($\tau_1 = 1.26 \pm 0.10$\,Gyr and $\tau_2 = 11.3 \pm 0.9$\,Gyr).

As seen from Fig. \ref{fig:spitoni}, the two-infall model explains the main
features of the abundance-age relations, but there are 
problems with the details. In the upper panel, the model predicts too
high ages for the old sequence. 
This part of the fit is improved if the model begins with 
$\feh = -1$ at an age of 11\,Gyr instead of 12\,Gyr, but then the fit to the 
steep decline of \ofe\ and \mgfe\ at high ages is less satisfactory. 
Furthermore, the model does not reach the group of 12 metal-rich stars 
on the old sequence having $\ofe \sim -0.1$ and a too large bump in \mgfe\
is predicted at ages between 6 and 8\,Gyr. Finally the model predicts a 
too steep decline of \mgfe\ at ages below 5\,Gyr.

It is beyond the scope of this paper to investigate if the fitting of the 
two-infall model to the data can be improved by varying the yields or by changing
assumptions about the star-formation history, the IMF, the time-scale
of Type\,Ia SNe, and the composition of the infalling gas. A comparison with
the age trends of other element ratios, such as \nafe , \nife , and \ymg ,
is also postponed. Instead, we briefly discuss if the alternative
scenario of forming the thick and the thin disks proposed by \citet{haywood19} can
explain the two sequences in the age-metallicity diagram and the corresponding age
trends of element ratios.

According to  \citet{haywood19}, high-alpha stars started to form in a
turbulent gaseous thick disk with a chemical evolution 
according to the closed-box model. After  3-4\,Gyr, the gas had been enriched
to solar metallicity but then experienced a quenching of star formation possibly
caused by formation of a central bar \citep{haywood18}. At the same time, accretion
of metal-poor gas mixed with solar-metallicity gas from the
thick disk began in the outer parts of the disk leading to the 
formation of the thin disk. In the solar neighbourhood, this caused a decrease
of metallicity to $\feh \sim -0.2$, but farther out in the disk a lower metallicity
was reached because of a higher portion of metal-poor gas. In this way, a radial
metallicity gradient in the thin disk was created. After the quenching period, the
chemical evolution continued leading to the formation of metal-rich stars
($\feh \sim +0.3$) in the inner disk (galactocentric distances $R_G < 7$\,kpc) 
and a gradual increase
of \feh\ with age in the outer disk \citep[see Fig. 6 in][for a sketch of the 
resulting age trends of \feh\ and \alphafe]{haywood19}. The trends
are similar to the predictions of the two-infall model shown in Fig. \ref{fig:spitoni},
but with a turnover in \feh\ at solar metallicity 
instead of $\feh = +0.3$ probably leading to a smaller bump in \mgfe\ at ages
between 6 and 8\,Gyr in better agreement with the data. The metal-rich
group of stars with low values of \ofe\ could then be interpreted as formed at  
the end of the inner disk evolution. It remains, however, to be
seen how well a detailed chemical evolution model for this scenario 
can explain the data in Fig. \ref{fig:spitoni}.

\begin{figure}
\resizebox{\hsize}{!}{\includegraphics{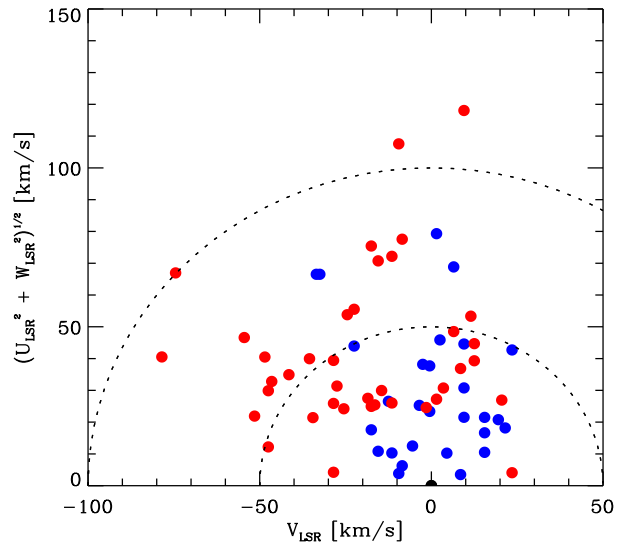}}
\caption{Toomre diagram for stars on the two age sequences 
in Fig. \ref{fig:feh-age}. Stars on the old age sequence are shown with 
filled red circles and those on the younger age sequence with filled blue circles.
The dotted lines correspond to respectively $V_{\rm total} = 50$\,\kmprs\ and 
$V_{\rm total} = 100$\,\kmprs . The uncertainty of the velocity components
is about 1.5\,\kmprs , corresponding to the size of the circles.}
\label{fig:toomre}
\end{figure}

In connection with the \citet{haywood19} scenario, the kinematics of 
our stars is of interest. Galactic velocity components,
$U, V, W$, and peri- and apo-galactic orbital distances from the 
Galactic centre, $R_{\rm peri}$ and $R_{\rm apo}$,
were adopted from \citet{holmberg09}, who used Tycho-2
proper motions \citep{hoeg00}, Hipparcos distances \citep{vanleeuwen07},
and radial velocities from the Geneva-Copenhagen survey \citep{nordstrom04} 
in their calculations. As seen from the resulting Toomre diagram  
in Fig. \ref{fig:toomre}, the old (red) population has a larger velocity dispersion
and a larger rotational lag with respect to the LSR
than the young (blue) population. 
This difference in kinematics is also evident from  the mean 
galactocentric distances in the stellar orbits
$R_m = (R_{\rm peri} + R_{\rm apo}) / 2$. The old population has an average
$\langle R_m \rangle = 7.34$\,kpc with an rms dispersion of $\sigma = 0.71$\,kpc. 
The younger population has $\langle R_m \rangle = 7.82$\,kpc and $\sigma = 0.44$\,kpc. 
Assuming that $R_{\rm m}$ is a measure of
the galactocentric distance of the stellar birthplace \citep[e.g.][]{edvardsson93}
these values support that the old stars have been formed inside the 
mean galactocentric distance for the formation of the younger population,
but the difference in $\langle R_m \rangle$ is small, so the kinematics 
does not provide much evidence of the  \citet{haywood19} disk
formation scenario. If, on the other hand, the orbital angular momenta
of the old stars have been changed, we cannot use $R_{\rm m}$ as a proxy
of stellar birthplace and therefore the stars may have been formed farther away.
 
\section{Summary and conclusions}
\label{conclusions}
High-precision chemical abundances have been determined
from HARPS spectra of 72 nearby solar-type stars and precise ages 
were derived by comparing spectroscopic effective temperatures
and luminosities based on {\em Gaia} DR2 distances to ASTEC isochrones. These data 
suggest that there are two distinct sequences in the age-metallicity diagram
(Fig.  \ref{fig:feh-age}). A similar distribution  is
seen in the age-metallicity diagram of dwarf stars for which precise
asteroseismic ages have been determined from {\em Kepler} satellite observations
of oscillation frequencies (Fig. \ref{fig:victor}). The two sequences can also be seen as 
distinct \xfe -age trends for several elements, most notable O, Na, Ca, and Ni
(Fig. \ref{fig:xfe-age}). 
 
Previous studies of the age-metallicity relation for stars in the solar
neighbourhood have not revealed any bimodal distribution. This  can be ascribed to
typical age uncertainties of 2-4\,Gyr, which blurs the age gap 
between the two sequences leading to an apparently flat distribution of \feh\ as
a function of age. The corresponding large scatter in \feh\ at a given age has 
been explained as due to migration of stars over several kpc
in a Galactic disk with a radial metallicity gradient and this has been the
most important argument for the existence of `churning' (changes of orbital angular
momentum) in addition to `blurring' (increase in epicycle amplitude in time).
With two distinct sequences in the age-metallicity diagram,
the scatter in \feh\ at a given age for each sequence is decreased
and it is a question if `churning' is needed to explain the data.

The elemental abundances provided in this paper have been derived by analysing
HARPS spectra with 1D model atmospheres under the LTE assumption. As discussed
in Sect. \ref{results}, a 3D non-LTE analysis may lead to changes of the derived
abundances and atmospheric parameters and therefore also the derived ages, although
the cases of C and O, for which 3D non-LTE corrections are available, suggest
that differential corrections are small for the lines and stars considered
in the present work. Still, it would be important to
extend 3D non-LTE calculations to other elements than C and O,
in particular to the \FeI\ lines used to derive \teff\ and \logg .
3D non-LTE corrections may also affect the derived jumps in some abundance ratios
between metal-rich stars on the old age sequence
and metal-poor stars on the younger sequence for ages at 6-7\,Gyr,
and therefore be important when testing chemical evolution models.

As discussed in Sect. \ref{discussion}, the two-infall chemical evolution
model \citep{chiappini97, spitoni19} provides an overall satisfactory
fit to the distribution of stars in the \fehi - , \ofe -, and \mgfe - age diagrams,
but some details such as the existence of a group of metal-rich stars 
with $\ofe \sim -0.10$ cannot be explained. It also remains to be investigated
if other abundance trends, in particular \ymg\ versus age, can be explained.
\ymg\ shows a remarkable steep and tight trend as function of age (except for
a puzzling deviation of the visual binary star $\zeta$\,Ret) with no
offset between the two age sequences and with negligible dependence on
metallicity in the range $-0.3 < \feh < +0.3$.
 
As an alternative to the two-infall model it should be investigated
if the scenario proposed by \citet{haywood19}, in which the inner thick disk 
evolves according to the closed box model and the outer thin disk forms from 
infalling metal-poor gas, can explain the data. Given that the Sun is 
situated at the transition between the two regions, the solar neighbourhood 
may consist of a mixture of stars having experienced two different 
chemical evolution tracks.

Given the relative small sample of stars studied,
it would be important to investigate if larger samples with
high precision abundances and ages confirm the split of the age-metallicity 
diagram into two age sequences. It requires, however, spectra of 
a similar high quality as the HARPS spectra applied in this paper to determine
abundances and isochrone ages with sufficient precision to detect the
gap between the two age sequences.  Another interesting possibility is to expand the
sample of solar-type stars with seismic ages derived from individual
oscillation frequencies using for example {\em TESS} satellite data.

\begin{acknowledgements}
We thank Anish Amarsi and Bengt Gustafsson for helpful comments on a first 
version of the manuscript and the referee for a very 
constructive report on the paper. 
Funding for the Stellar Astrophysics Centre is provided by The
Danish National Research Foundation (Grant agreement no.: DNRF106).
V.S.A. acknowledges support from the Independent Research Fund Denmark
(Research grant 7027-00096B). V.S.A and J.R.M acknowledge support 
from the Carlsberg foundation (grant agreement CF19-0649).
This research has made use of the SIMBAD database,
operated at CDS, Strasbourg, France, as well as
data from the European Space Agency (ESA)
mission {\em Gaia} (\url{https://www.cosmos.esa.int/gaia}), processed by
the {\it Gaia} Data Processing and Analysis Consortium (DPAC,
\url{https://www.cosmos.esa.int/web/gaia/dpac/consortium}). Funding
for the DPAC has been provided by national institutions, in particular
the institutions participating in the {\it Gaia} Multilateral Agreement.
Based on observations collected at the European Southern Observatory
under ESO HARPS programs
060.A-9036,
072.C-0488,
072.C-0513,
073.D-0578,
074.C-0012,
076.C-0878,
077.C-0530,
078.C-0833,
078.C-0044,
079.C-0681,
085.C-0019,
089.C-0732,
091.C-0936,
092.C-0721,
093.C-0409,
093.C-0919,
094.C-0901,
097,C-0571,
100.D-0444,
101.C-0275,
102.C-0584,
183.C-0972,
183.D-0729,
188.C-0265,
192.C-0224,
192.C-0852,
196.C-1006,
198.C-0836,
and ESO FEROS programs
092.A-9002,
095.A-9029,
099.A-9022,
100.A-9022.

\end{acknowledgements}

\bibliographystyle{aa}
\bibliography{nissen.2020}

\begin{thebibliography}{130}
\expandafter\ifx\csname natexlab\endcsname\relax\def\natexlab#1{#1}\fi

\bibitem[{{Adibekyan} {et~al.}(2016){Adibekyan}, {Delgado-Mena}, {Figueira},
  {Sousa}, {Santos}, {Faria}, {Gonz{\'a}lez Hern{\'a}ndez}, {Israelian},
  {Harutyunyan}, {Su{\'a}rez-Andr{\'e}s}, \& {Hakobyan}}]{adibekyan16}
{Adibekyan}, V., {Delgado-Mena}, E., {Figueira}, P., {et~al.} 2016, \aap, 591,
  A34

\bibitem[{{Adibekyan} {et~al.}(2012){Adibekyan}, {Sousa}, {Santos}, {Delgado
  Mena}, {Gonz{\'a}lez Hern{\'a}ndez}, {Israelian}, {Mayor}, \&
  {Khachatryan}}]{adibekyan12}
{Adibekyan}, V.~Z., {Sousa}, S.~G., {Santos}, N.~C., {et~al.} 2012, \aap, 545,
  A32

\bibitem[{{Allende Prieto} {et~al.}(2002){Allende Prieto}, {Asplund},
  {Garc{\'\i}a L{\'o}pez}, \& {Lambert}}]{allende-prieto02}
{Allende Prieto}, C., {Asplund}, M., {Garc{\'\i}a L{\'o}pez}, R.~J., \&
  {Lambert}, D.~L. 2002, \apj, 567, 544

\bibitem[{{Allende Prieto} {et~al.}(2001){Allende Prieto}, {Lambert}, \&
  {Asplund}}]{allende-prieto01}
{Allende Prieto}, C., {Lambert}, D.~L., \& {Asplund}, M. 2001, \apjl, 556, L63

\bibitem[{{Amarsi} {et~al.}(2018){Amarsi}, {Barklem}, {Asplund}, {Collet}, \&
  {Zatsarinny}}]{amarsi18}
{Amarsi}, A.~M., {Barklem}, P.~S., {Asplund}, M., {Collet}, R., \&
  {Zatsarinny}, O. 2018, \aap, 616, A89

\bibitem[{{Amarsi} {et~al.}(2019{\natexlab{a}}){Amarsi}, {Barklem}, {Collet},
  {Grevesse}, \& {Asplund}}]{amarsi19a}
{Amarsi}, A.~M., {Barklem}, P.~S., {Collet}, R., {Grevesse}, N., \& {Asplund},
  M. 2019{\natexlab{a}}, \aap, 624, A111

\bibitem[{{Amarsi} {et~al.}(2019{\natexlab{b}}){Amarsi}, {Nissen}, \&
  {Sk{\'u}lad{\'o}ttir}}]{amarsi19b}
{Amarsi}, A.~M., {Nissen}, P.~E., \& {Sk{\'u}lad{\'o}ttir}, {\'A}.
  2019{\natexlab{b}}, \aap, 630, A104

\bibitem[{{Anders} {et~al.}(2017){Anders}, {Chiappini}, {Minchev}, {Miglio},
  {Montalb{\'a}n}, {Mosser}, {Rodrigues}, {Santiago}, {Baudin}, {Beers}, {da
  Costa}, {Garc{\'\i}a}, {Garc{\'\i}a-Hern{\'a}ndez}, {Holtzman}, {Maia},
  {Majewski}, {Mathur}, {Noels-Grotsch}, {Pan}, {Schneider}, {Schultheis},
  {Steinmetz}, {Valentini}, \& {Zamora}}]{anders17}
{Anders}, F., {Chiappini}, C., {Minchev}, I., {et~al.} 2017, \aap, 600, A70

\bibitem[{{Asplund} {et~al.}(2009){Asplund}, {Grevesse}, {Sauval}, \&
  {Scott}}]{asplund09}
{Asplund}, M., {Grevesse}, N., {Sauval}, A.~J., \& {Scott}, P. 2009, \araa, 47,
  481

\bibitem[{{Asplund} {et~al.}(2000){Asplund}, {Nordlund}, {Trampedach}, {Allende
  Prieto}, \& {Stein}}]{asplund00}
{Asplund}, M., {Nordlund}, {\r{A}}., {Trampedach}, R., {Allende Prieto}, C., \&
  {Stein}, R.~F. 2000, \aap, 359, 729

\bibitem[{{Barklem}(2016)}]{barklem16}
{Barklem}, P.~S. 2016, \aapr, 24, 9

\bibitem[{{Basu} \& {Antia}(2004)}]{basu04}
{Basu}, S. \& {Antia}, H.~M. 2004, \apjl, 606, L85

\bibitem[{{Bedell} {et~al.}(2018){Bedell}, {Bean}, {Mel{\'e}ndez}, {Spina},
  {Ram{\'\i}rez}, {Asplund}, {Alves-Brito}, {dos Santos}, {Dreizler}, {Yong},
  {Monroe}, \& {Casagrande}}]{bedell18}
{Bedell}, M., {Bean}, J.~L., {Mel{\'e}ndez}, J., {et~al.} 2018, \apj, 865, 68

\bibitem[{{Bensby} {et~al.}(2005){Bensby}, {Feltzing}, {Lundstr{\"o}m}, \&
  {Ilyin}}]{bensby05}
{Bensby}, T., {Feltzing}, S., {Lundstr{\"o}m}, I., \& {Ilyin}, I. 2005, \aap,
  433, 185

\bibitem[{{Bensby} {et~al.}(2014){Bensby}, {Feltzing}, \& {Oey}}]{bensby14}
{Bensby}, T., {Feltzing}, S., \& {Oey}, M.~S. 2014, \aap, 562, A71

\bibitem[{{Bond} {et~al.}(2010){Bond}, {O'Brien}, \& {Lauretta}}]{bond10}
{Bond}, J.~C., {O'Brien}, D.~P., \& {Lauretta}, D.~S. 2010, \apj, 715, 1050

\bibitem[{{Brewer} \& {Fischer}(2016)}]{brewer16}
{Brewer}, J.~M. \& {Fischer}, D.~A. 2016, \apj, 831, 20

\bibitem[{{Buck}(2020)}]{buck20}
{Buck}, T. 2020, \mnras, 491, 5435

\bibitem[{{Buder} {et~al.}(2019){Buder}, {Lind}, {Ness}, {Asplund}, {Duong},
  {Lin}, {Kos}, {Casagrande}, {Casey}, {Bland-Hawthorn}, {de Silva}, {D'Orazi},
  {Freeman}, {Martell}, {Schlesinger}, {Sharma}, {Simpson}, {Zucker},
  {Zwitter}, {{\v{C}}otar}, {Dotter}, {Hayden}, {Hyde}, {Kafle}, {Lewis},
  {Nataf}, {Nordlander}, {Reid}, {Rix}, {Sk{\'u}lad{\'o}ttir}, {Stello},
  {Ting}, {Traven}, {Wyse}, \& {Galah Collaboration}}]{buder19}
{Buder}, S., {Lind}, K., {Ness}, M.~K., {et~al.} 2019, \aap, 624, A19

\bibitem[{{Casagrande} {et~al.}(2010){Casagrande}, {Ram{\'{\i}}rez},
  {Mel{\'e}ndez}, {Bessell}, \& {Asplund}}]{casagrande10}
{Casagrande}, L., {Ram{\'{\i}}rez}, I., {Mel{\'e}ndez}, J., {Bessell}, M., \&
  {Asplund}, M. 2010, \aap, 512, A54

\bibitem[{{Casagrande} {et~al.}(2011){Casagrande}, {Sch{\"o}nrich}, {Asplund},
  {Cassisi}, {Ram{\'\i}rez}, {Mel{\'e}ndez}, {Bensby}, \&
  {Feltzing}}]{casagrande11}
{Casagrande}, L., {Sch{\"o}nrich}, R., {Asplund}, M., {et~al.} 2011, \aap, 530,
  A138

\bibitem[{{Chambers}(2010)}]{chambers10}
{Chambers}, J.~E. 2010, \apj, 724, 92

\bibitem[{{Chiappini} {et~al.}(1997){Chiappini}, {Matteucci}, \&
  {Gratton}}]{chiappini97}
{Chiappini}, C., {Matteucci}, F., \& {Gratton}, R. 1997, \apj, 477, 765

\bibitem[{{Christensen-Dalsgaard}(2008)}]{jcd08a}
{Christensen-Dalsgaard}, J. 2008, \apss, 316, 13

\bibitem[{{Christensen-Dalsgaard} \& {P{\'e}rez Hern{\'a}ndez}(1991)}]{jcd91}
{Christensen-Dalsgaard}, J. \& {P{\'e}rez Hern{\'a}ndez}, F. 1991, in Lecture
  Notes in Physics, Berlin Springer Verlag, Vol. 388, Challenges to Theories of
  the Structure of Moderate-Mass Stars, ed. D.~{Gough} \& J.~{Toomre}, 43

\bibitem[{{Cutri} {et~al.}(2003){Cutri}, {Skrutskie}, {van Dyk}, {Beichman},
  {Carpenter}, {Chester}, {Cambresy}, {Evans}, {Fowler}, {Gizis}, {Howard},
  {Huchra}, {Jarrett}, {Kopan}, {Kirkpatrick}, {Light}, {Marsh}, {McCallon},
  {Schneider}, {Stiening}, {Sykes}, {Weinberg}, {Wheaton}, {Wheelock}, \&
  {Zacarias}}]{cutri03}
{Cutri}, R.~M., {Skrutskie}, M.~F., {van Dyk}, S., {et~al.} 2003, VizieR Online
  Data Catalog, II/246

\bibitem[{{da Silva} {et~al.}(2012){da Silva}, {Porto de Mello}, {Milone}, {da
  Silva}, {Ribeiro}, \& {Rocha-Pinto}}]{dasilva12}
{da Silva}, R., {Porto de Mello}, G.~F., {Milone}, A.~C., {et~al.} 2012, \aap,
  542, A84

\bibitem[{{Delgado Mena} {et~al.}(2010){Delgado Mena}, {Israelian},
  {Gonz{\'a}lez Hern{\'a}ndez}, {Bond}, {Santos}, {Udry}, \&
  {Mayor}}]{delgado-mena10}
{Delgado Mena}, E., {Israelian}, G., {Gonz{\'a}lez Hern{\'a}ndez}, J.~I.,
  {et~al.} 2010, \apj, 725, 2349

\bibitem[{{Delgado Mena} {et~al.}(2019){Delgado Mena}, {Moya}, {Adibekyan},
  {Tsantaki}, {Gonz{\'a}lez Hern{\'a}ndez}, {Israelian}, {Davies}, {Chaplin},
  {Sousa}, {Ferreira}, \& {Santos}}]{delgado-mena19}
{Delgado Mena}, E., {Moya}, A., {Adibekyan}, V., {et~al.} 2019, \aap, 624, A78

\bibitem[{{Edvardsson} {et~al.}(1993){Edvardsson}, {Andersen}, {Gustafsson},
  {Lambert}, {Nissen}, \& {Tomkin}}]{edvardsson93}
{Edvardsson}, B., {Andersen}, J., {Gustafsson}, B., {et~al.} 1993, \aap, 275,
  101

\bibitem[{{Eiroa} {et~al.}(2010){Eiroa}, {Fedele}, {Maldonado},
  {Gonz{\'a}lez-Garc{\'\i}a}, {Rodmann}, {Heras}, {Pilbratt}, {Augereau},
  {Mora}, {Montesinos}, {Ardila}, {Bryden}, {Liseau}, {Stapelfeldt},
  {Launhardt}, {Solano}, {Bayo}, {Absil}, {Ar{\'e}valo}, {Barrado},
  {Beichmann}, {Danchi}, {Del Burgo}, {Ertel}, {Fridlund}, {Fukagawa},
  {Guti{\'e}rrez}, {Gr{\"u}n}, {Kamp}, {Krivov}, {Lebreton}, {L{\"o}hne},
  {Lorente}, {Marshall}, {Mart{\'\i}nez-Arn{\'a}iz}, {Meeus}, {Montes},
  {Morbidelli}, {M{\"u}ller}, {Mutschke}, {Nakagawa}, {Olofsson}, {Ribas},
  {Roberge}, {Sanz-Forcada}, {Th{\'e}bault}, {Walker}, {White}, \&
  {Wolf}}]{Eiroa10}
{Eiroa}, C., {Fedele}, D., {Maldonado}, J., {et~al.} 2010, \aap, 518, L131

\bibitem[{{Epstein} {et~al.}(2010){Epstein}, {Johnson}, {Dong}, {Udalski},
  {Gould}, \& {Becker}}]{epstein10}
{Epstein}, C.~R., {Johnson}, J.~A., {Dong}, S., {et~al.} 2010, \apj, 709, 447

\bibitem[{{Faramaz} {et~al.}(2018){Faramaz}, {Bryden}, {Stapelfeldt}, {Booth},
  {Bayo}, {Beust}, {Casassus}, {Cuadra}, {Hales}, {Hughes}, {Olofsson}, {Su},
  \& {Wilner}}]{faramaz18}
{Faramaz}, V., {Bryden}, G., {Stapelfeldt}, K.~R., {et~al.} 2018, \mnras, 481,
  44

\bibitem[{{Feltzing} {et~al.}(2001){Feltzing}, {Holmberg}, \&
  {Hurley}}]{Feltzing01}
{Feltzing}, S., {Holmberg}, J., \& {Hurley}, J.~R. 2001, \aap, 377, 911

\bibitem[{{Feltzing} {et~al.}(2017){Feltzing}, {Howes}, {McMillan}, \&
  {Stonkut{\.e}}}]{feltzing17}
{Feltzing}, S., {Howes}, L.~M., {McMillan}, P.~J., \& {Stonkut{\.e}}, E. 2017,
  \mnras, 465, L109

\bibitem[{{Feuillet} {et~al.}(2019){Feuillet}, {Frankel}, {Lind}, {Frinchaboy},
  {Garc{\'\i}a-Hern{\'a}ndez}, {Lane}, {Nitschelm}, \&
  {Roman-Lopes}}]{feuillet19}
{Feuillet}, D.~K., {Frankel}, N., {Lind}, K., {et~al.} 2019, \mnras, 489, 1742

\bibitem[{{Flores} {et~al.}(2018){Flores}, {Saffe}, {Buccino}, {Jaque
  Arancibia}, {Gonz{\'a}lez}, {Nu{\~n}ez}, \& {Jofr{\'e}}}]{flores18}
{Flores}, M., {Saffe}, C., {Buccino}, A., {et~al.} 2018, \mnras, 476, 2751

\bibitem[{{Frankel} {et~al.}(2018){Frankel}, {Rix}, {Ting}, {Ness}, \&
  {Hogg}}]{frankel18}
{Frankel}, N., {Rix}, H.-W., {Ting}, Y.-S., {Ness}, M., \& {Hogg}, D.~W. 2018,
  \apj, 865, 96

\bibitem[{{Fuhrmann}(1998)}]{fuhrmann98}
{Fuhrmann}, K. 1998, \aap, 338, 161

\bibitem[{{Fuhrmann} {et~al.}(2017){Fuhrmann}, {Chini}, {Kaderhandt}, \&
  {Chen}}]{fuhrmann17}
{Fuhrmann}, K., {Chini}, R., {Kaderhandt}, L., \& {Chen}, Z. 2017, \mnras, 464,
  2610

\bibitem[{{Gaia Collaboration} {et~al.}(2018){Gaia Collaboration}, {Brown},
  {Vallenari}, {Prusti}, {de Bruijne}, {Babusiaux}, {Bailer-Jones}, {Biermann},
  {Evans}, {Eyer}, \& et~al.}]{gaia.collaboration18}
{Gaia Collaboration}, {Brown}, A.~G.~A., {Vallenari}, A., {et~al.} 2018, \aap,
  616, A1

\bibitem[{{Gaidos}(2015)}]{gaidos15}
{Gaidos}, E. 2015, \apj, 804, 40

\bibitem[{{Grand} {et~al.}(2018){Grand}, {Bustamante}, {G{\'o}mez}, {Kawata},
  {Marinacci}, {Pakmor}, {Rix}, {Simpson}, {Sparre}, \& {Springel}}]{grand18}
{Grand}, R. J.~J., {Bustamante}, S., {G{\'o}mez}, F.~A., {et~al.} 2018, \mnras,
  474, 3629

\bibitem[{{Gratton} {et~al.}(2000){Gratton}, {Carretta}, {Matteucci}, \&
  {Sneden}}]{gratton00}
{Gratton}, R.~G., {Carretta}, E., {Matteucci}, F., \& {Sneden}, C. 2000, \aap,
  358, 671

\bibitem[{{Gray}(1992)}]{gray92}
{Gray}, D.~F. 1992, {The observation and analysis of stellar photospheres.
  Camb. Astrophys. Ser., 20 (Cambridge: Cambridge Univ. Press)}

\bibitem[{{Grevesse} {et~al.}(1996){Grevesse}, {Noels}, \&
  {Sauval}}]{grevesse96}
{Grevesse}, N., {Noels}, A., \& {Sauval}, A.~J. 1996, in Astronomical Society
  of the Pacific Conference Series, Vol.~99, Cosmic Abundances, ed. S.~S.
  {Holt} \& G.~{Sonneborn}, 117

\bibitem[{{Grevesse} {et~al.}(2015){Grevesse}, {Scott}, {Asplund}, \&
  {Sauval}}]{grevesse15}
{Grevesse}, N., {Scott}, P., {Asplund}, M., \& {Sauval}, A.~J. 2015, \aap, 573,
  A27

\bibitem[{{Grisoni} {et~al.}(2017){Grisoni}, {Spitoni}, {Matteucci},
  {Recio-Blanco}, {de Laverny}, {Hayden}, {Mikolaitis}, \&
  {Worley}}]{grisoni17}
{Grisoni}, V., {Spitoni}, E., {Matteucci}, F., {et~al.} 2017, \mnras, 472, 3637

\bibitem[{{Gustafsson}(2018{\natexlab{a}})}]{gustafsson18a}
{Gustafsson}, B. 2018{\natexlab{a}}, \aap, 616, A91

\bibitem[{{Gustafsson}(2018{\natexlab{b}})}]{gustafsson18b}
{Gustafsson}, B. 2018{\natexlab{b}}, \aap, 620, A53

\bibitem[{{Gustafsson} {et~al.}(2008){Gustafsson}, {Edvardsson}, {Eriksson},
  {J{\o}rgensen}, {Nordlund}, \& {Plez}}]{gustafsson08}
{Gustafsson}, B., {Edvardsson}, B., {Eriksson}, K., {et~al.} 2008, \aap, 486,
  951

\bibitem[{{Hayden} {et~al.}(2015){Hayden}, {Bovy}, {Holtzman}, {Nidever},
  {Bird}, {Weinberg}, {Andrews}, {Majewski}, {Allende Prieto}, {Anders},
  {Beers}, {Bizyaev}, {Chiappini}, {Cunha}, {Frinchaboy},
  {Garc{\'\i}a-Her{\'n}and ez}, {Garc{\'\i}a P{\'e}rez}, {Girardi}, {Harding},
  {Hearty}, {Johnson}, {M{\'e}sz{\'a}ros}, {Minchev}, {O'Connell}, {Pan},
  {Robin}, {Schiavon}, {Schneider}, {Schultheis}, {Shetrone}, {Skrutskie},
  {Steinmetz}, {Smith}, {Wilson}, {Zamora}, \& {Zasowski}}]{hayden15}
{Hayden}, M.~R., {Bovy}, J., {Holtzman}, J.~A., {et~al.} 2015, \apj, 808, 132

\bibitem[{{Haywood} {et~al.}(2018){Haywood}, {Di Matteo}, {Lehnert}, {Snaith},
  {Fragkoudi}, \& {Khoperskov}}]{haywood18}
{Haywood}, M., {Di Matteo}, P., {Lehnert}, M., {et~al.} 2018, \aap, 618, A78

\bibitem[{{Haywood} {et~al.}(2013){Haywood}, {Di Matteo}, {Lehnert}, {Katz}, \&
  {G{\'o}mez}}]{haywood13}
{Haywood}, M., {Di Matteo}, P., {Lehnert}, M.~D., {Katz}, D., \& {G{\'o}mez},
  A. 2013, \aap, 560, A109

\bibitem[{{Haywood} {et~al.}(2019){Haywood}, {Snaith}, {Lehnert}, {Di Matteo},
  \& {Khoperskov}}]{haywood19}
{Haywood}, M., {Snaith}, O., {Lehnert}, M.~D., {Di Matteo}, P., \&
  {Khoperskov}, S. 2019, \aap, 625, A105

\bibitem[{{H{\o}g} {et~al.}(2000){H{\o}g}, {Fabricius}, {Makarov}, {Urban},
  {Corbin}, {Wycoff}, {Bastian}, {Schwekendiek}, \& {Wicenec}}]{hoeg00}
{H{\o}g}, E., {Fabricius}, C., {Makarov}, V.~V., {et~al.} 2000, \aap, 355, L27

\bibitem[{{Holmberg} {et~al.}(2009){Holmberg}, {Nordstr{\"o}m}, \&
  {Andersen}}]{holmberg09}
{Holmberg}, J., {Nordstr{\"o}m}, B., \& {Andersen}, J. 2009, \aap, 501, 941

\bibitem[{{Huber} {et~al.}(2013){Huber}, {Chaplin}, {Christensen-Dalsgaard},
  {Gilliland}, {Kjeldsen}, {Buchhave}, {Fischer}, {Lissauer}, {Rowe},
  {Sanchis-Ojeda}, {Basu}, {Handberg}, {Hekker}, {Howard}, {Isaacson},
  {Karoff}, {Latham}, {Lund}, {Lundkvist}, {Marcy}, {Miglio}, {Silva Aguirre},
  {Stello}, {Arentoft}, {Barclay}, {Bedding}, {Burke}, {Christiansen},
  {Elsworth}, {Haas}, {Kawaler}, {Metcalfe}, {Mullally}, \&
  {Thompson}}]{huber13}
{Huber}, D., {Chaplin}, W.~J., {Christensen-Dalsgaard}, J., {et~al.} 2013,
  \apj, 767, 127

\bibitem[{{Jofr{\'e}} {et~al.}(2020){Jofr{\'e}}, {Jackson}, \& {Tucci
  Maia}}]{jofre20}
{Jofr{\'e}}, P., {Jackson}, H., \& {Tucci Maia}, M. 2020, \aap, 633, L9

\bibitem[{{Johansson} {et~al.}(2003){Johansson}, {Litz{\'e}n}, {Lundberg}, \&
  {Zhang}}]{johansson03}
{Johansson}, S., {Litz{\'e}n}, U., {Lundberg}, H., \& {Zhang}, Z. 2003, \apjl,
  584, L107

\bibitem[{{Karakas} \& {Lugaro}(2016)}]{karakas16}
{Karakas}, A.~I. \& {Lugaro}, M. 2016, \apj, 825, 26

\bibitem[{{Kim} {et~al.}(2002){Kim}, {Demarque}, {Yi}, \& {Alexander}}]{kim02}
{Kim}, Y.-C., {Demarque}, P., {Yi}, S.~K., \& {Alexander}, D.~R. 2002, \apjs,
  143, 499

\bibitem[{{Kuchner} \& {Seager}(2005)}]{kuchner05}
{Kuchner}, M.~J. \& {Seager}, S. 2005, ArXiv Astrophysics e-prints
  [\eprint{astro-ph/0504214}]

\bibitem[{{Lind} {et~al.}(2017){Lind}, {Amarsi}, {Asplund}, {Barklem},
  {Bautista}, {Bergemann}, {Collet}, {Kiselman}, {Leenaarts}, \&
  {Pereira}}]{lind17}
{Lind}, K., {Amarsi}, A.~M., {Asplund}, M., {et~al.} 2017, \mnras, 468, 4311

\bibitem[{{Lindegren} {et~al.}(2018){Lindegren}, {Hern{\'a}ndez}, {Bombrun},
  {Klioner}, {Bastian}, {Ramos-Lerate}, {de Torres}, {Steidelm{\"u}ller},
  {Stephenson}, {Hobbs}, {Lammers}, {Biermann}, {Geyer}, {Hilger}, {Michalik},
  {Stampa}, {McMillan}, {Casta{\~n}eda}, {Clotet}, {Comoretto}, {Davidson},
  {Fabricius}, {Gracia}, {Hambly}, {Hutton}, {Mora}, {Portell}, {van Leeuwen},
  {Abbas}, {Abreu}, {Altmann}, {Andrei}, {Anglada}, {Balaguer-N{\'u}{\~n}ez},
  {Barache}, {Becciani}, {Bertone}, {Bianchi}, {Bouquillon}, {Bourda},
  {Br{\"u}semeister}, {Bucciarelli}, {Busonero}, {Buzzi}, {Cancelliere},
  {Carlucci}, {Charlot}, {Cheek}, {Crosta}, {Crowley}, {de Bruijne}, {de
  Felice}, {Drimmel}, {Esquej}, {Fienga}, {Fraile}, {Gai}, {Garralda},
  {Gonz{\'a}lez-Vidal}, {Guerra}, {Hauser}, {Hofmann}, {Holl}, {Jordan},
  {Lattanzi}, {Lenhardt}, {Liao}, {Licata}, {Lister}, {L{\"o}ffler},
  {Marchant}, {Martin-Fleitas}, {Messineo}, {Mignard}, {Morbidelli}, {Poggio},
  {Riva}, {Rowell}, {Salguero}, {Sarasso}, {Sciacca}, {Siddiqui}, {Smart},
  {Spagna}, {Steele}, {Taris}, {Torra}, {van Elteren}, {van Reeven}, \&
  {Vecchiato}}]{lindegren18}
{Lindegren}, L., {Hern{\'a}ndez}, J., {Bombrun}, A., {et~al.} 2018, \aap, 616,
  A2

\bibitem[{{Lodders}(2003)}]{Lodders03}
{Lodders}, K. 2003, \apj, 591, 1220

\bibitem[{{Ludwig} \& {Steffen}(2016)}]{ludwig16}
{Ludwig}, H.~G. \& {Steffen}, M. 2016, Astronomische Nachrichten, 337, 844

\bibitem[{{Lund} {et~al.}(2017){Lund}, {Silva Aguirre}, {Davies}, {Chaplin},
  {Christensen-Dalsgaard}, {Houdek}, {White}, {Bedding}, {Ball}, {Huber},
  {Antia}, {Lebreton}, {Latham}, {Handberg}, {Verma}, {Basu}, {Casagrande},
  {Justesen}, {Kjeldsen}, \& {Mosumgaard}}]{lund17}
{Lund}, M.~N., {Silva Aguirre}, V., {Davies}, G.~R., {et~al.} 2017, \apj, 835,
  172

\bibitem[{{Mamajek} \& {Hillenbrand}(2008)}]{mamajek08}
{Mamajek}, E.~E. \& {Hillenbrand}, L.~A. 2008, \apj, 687, 1264

\bibitem[{{Mayor} {et~al.}(2003){Mayor}, {Pepe}, {Queloz}, {Bouchy},
  {Rupprecht}, {Lo Curto}, {Avila}, {Benz}, {Bertaux}, {Bonfils}, {Dall},
  {Dekker}, {Delabre}, {Eckert}, {Fleury}, {Gilliotte}, {Gojak}, {Guzman},
  {Kohler}, {Lizon}, {Longinotti}, {Lovis}, {Megevand}, {Pasquini}, {Reyes},
  {Sivan}, {Sosnowska}, {Soto}, {Udry}, {van Kesteren}, {Weber}, \&
  {Weilenmann}}]{mayor03}
{Mayor}, M., {Pepe}, F., {Queloz}, D., {et~al.} 2003, The Messenger, 114, 20

\bibitem[{{Mel{\'e}ndez} {et~al.}(2009){Mel{\'e}ndez}, {Asplund}, {Gustafsson},
  \& {Yong}}]{melendez09}
{Mel{\'e}ndez}, J., {Asplund}, M., {Gustafsson}, B., \& {Yong}, D. 2009, \apjl,
  704, L66

\bibitem[{{Mel{\'e}ndez} {et~al.}(2017){Mel{\'e}ndez}, {Bedell}, {Bean},
  {Ram{\'\i}rez}, {Asplund}, {Dreizler}, {Yan}, {Shi}, {Lind}, {Ferraz-Mello},
  {Galarza}, {dos Santos}, {Spina}, {Maia}, {Alves-Brito}, {Monroe}, \&
  {Casagrande}}]{melendez17}
{Mel{\'e}ndez}, J., {Bedell}, M., {Bean}, J.~L., {et~al.} 2017, \aap, 597, A34

\bibitem[{{Michaud} \& {Proffitt}(1993)}]{michaud93}
{Michaud}, G. \& {Proffitt}, C.~R. 1993, Astronomical Society of the Pacific
  Conference Series, Vol.~40, {Particle transport processes.}, ed. W.~W.
  {Weiss} \& A.~{Baglin}, 246--259

\bibitem[{{Miglio} {et~al.}(2020){Miglio}, {Chiappini}, {Mackereth}, {Davies},
  {Brogaard}, {Casagrande}, {Chaplin}, {Girardi}, {Kawata}, {Khan}, {Izzard},
  {Montalban}, {Mosser}, {Vincenzo}, {Bossini}, {Noels}, {Rodrigues},
  {Valentini}, \& {Mand el}}]{miglio20}
{Miglio}, A., {Chiappini}, C., {Mackereth}, T., {et~al.} 2020, arXiv e-prints,
  arXiv:2004.14806

\bibitem[{{Minchev} {et~al.}(2013){Minchev}, {Chiappini}, \&
  {Martig}}]{minchev13}
{Minchev}, I., {Chiappini}, C., \& {Martig}, M. 2013, \aap, 558, A9

\bibitem[{{Mor} {et~al.}(2019){Mor}, {Robin}, {Figueras}, {Roca-F{\`a}brega},
  \& {Luri}}]{mor19}
{Mor}, R., {Robin}, A.~C., {Figueras}, F., {Roca-F{\`a}brega}, S., \& {Luri},
  X. 2019, \aap, 624, L1

\bibitem[{{Nagar} {et~al.}(2020){Nagar}, {Spina}, \& {Karakas}}]{nagar20}
{Nagar}, T., {Spina}, L., \& {Karakas}, A.~I. 2020, \apjl, 888, L9

\bibitem[{{Nidever} {et~al.}(2014){Nidever}, {Bovy}, {Bird}, {Andrews},
  {Hayden}, {Holtzman}, {Majewski}, {Smith}, {Robin}, {Garc{\'\i}a P{\'e}rez},
  {Cunha}, {Allende Prieto}, {Zasowski}, {Schiavon}, {Johnson}, {Weinberg},
  {Feuillet}, {Schneider}, {Shetrone}, {Sobeck}, {Garc{\'\i}a-Hern{\'a}ndez},
  {Zamora}, {Rix}, {Beers}, {Wilson}, {O'Connell}, {Minchev}, {Chiappini},
  {Anders}, {Bizyaev}, {Brewington}, {Ebelke}, {Frinchaboy}, {Ge}, {Kinemuchi},
  {Malanushenko}, {Malanushenko}, {Marchante}, {M{\'e}sz{\'a}ros}, {Oravetz},
  {Pan}, {Simmons}, \& {Skrutskie}}]{nidever14}
{Nidever}, D.~L., {Bovy}, J., {Bird}, J.~C., {et~al.} 2014, \apj, 796, 38

\bibitem[{{Nissen}(2013)}]{nissen13}
{Nissen}, P.~E. 2013, \aap, 552, A73

\bibitem[{{Nissen}(2015)}]{nissen15}
{Nissen}, P.~E. 2015, \aap, 579, A52 (Paper I)

\bibitem[{{Nissen}(2016)}]{nissen16}
{Nissen}, P.~E. 2016, \aap, 593, A65 (Paper II)

\bibitem[{{Nissen} {et~al.}(2014){Nissen}, {Chen}, {Carigi}, {Schuster}, \&
  {Zhao}}]{nissen14}
{Nissen}, P.~E., {Chen}, Y.~Q., {Carigi}, L., {Schuster}, W.~J., \& {Zhao}, G.
  2014, \aap, 568, A25

\bibitem[{{Nissen} \& {Gustafsson}(2018)}]{nissen18}
{Nissen}, P.~E. \& {Gustafsson}, B. 2018, \aapr, 26, 6

\bibitem[{{Nissen} \& {Schuster}(2010)}]{nissen10}
{Nissen}, P.~E. \& {Schuster}, W.~J. 2010, \aap, 511, L10

\bibitem[{{Nissen} {et~al.}(2017){Nissen}, {Silva Aguirre},
  {Christensen-Dalsgaard}, {Collet}, {Grundahl}, \& {Slumstrup}}]{nissen17}
{Nissen}, P.~E., {Silva Aguirre}, V., {Christensen-Dalsgaard}, J., {et~al.}
  2017, \aap, 608, A112 (Paper III)

\bibitem[{{Noguchi}(2018)}]{noguchi18}
{Noguchi}, M. 2018, \nat, 559, 585

\bibitem[{{Nordlander} \& {Lind}(2017)}]{nordlander17}
{Nordlander}, T. \& {Lind}, K. 2017, \aap, 607, A75

\bibitem[{{Nordstr{\"o}m} {et~al.}(2004){Nordstr{\"o}m}, {Mayor}, {Andersen},
  {Holmberg}, {Pont}, {J{\o}rgensen}, {Olsen}, {Udry}, \&
  {Mowlavi}}]{nordstrom04}
{Nordstr{\"o}m}, B., {Mayor}, M., {Andersen}, J., {et~al.} 2004, \aap, 418, 989

\bibitem[{{Oh} {et~al.}(2018){Oh}, {Price-Whelan}, {Brewer}, {Hogg}, {Spergel},
  \& {Myles}}]{oh18}
{Oh}, S., {Price-Whelan}, A.~M., {Brewer}, J.~M., {et~al.} 2018, \apj, 854, 138

\bibitem[{{Olsen}(1983)}]{olsen83}
{Olsen}, E.~H. 1983, \aaps, 54, 55

\bibitem[{{Osorio} \& {Barklem}(2016)}]{osorio16}
{Osorio}, Y. \& {Barklem}, P.~S. 2016, \aap, 586, A120

\bibitem[{{Pagel}(1997)}]{pagel97}
{Pagel}, B. E.~J. 1997, {Nucleosynthesis and Chemical Evolution of Galaxies.
  (Cambridge: Cambridge Univ. Press)}

\bibitem[{{Petigura} \& {Marcy}(2011)}]{petigura11}
{Petigura}, E.~A. \& {Marcy}, G.~W. 2011, \apj, 735, 41

\bibitem[{{Prochaska} {et~al.}(2000){Prochaska}, {Naumov}, {Carney},
  {McWilliam}, \& {Wolfe}}]{prochaska00}
{Prochaska}, J.~X., {Naumov}, S.~O., {Carney}, B.~W., {McWilliam}, A., \&
  {Wolfe}, A.~M. 2000, \aj, 120, 2513

\bibitem[{{Ram{\'\i}rez} {et~al.}(2019){Ram{\'\i}rez}, {Khanal}, {Lichon},
  {Chanam{\'e}}, {Endl}, {Mel{\'e}ndez}, \& {Lambert}}]{ramirez19}
{Ram{\'\i}rez}, I., {Khanal}, S., {Lichon}, S.~J., {et~al.} 2019, \mnras, 490,
  2448

\bibitem[{{Reddy} {et~al.}(2006){Reddy}, {Lambert}, \& {Allende
  Prieto}}]{reddy06}
{Reddy}, B.~E., {Lambert}, D.~L., \& {Allende Prieto}, C. 2006, \mnras, 367,
  1329

\bibitem[{{Rocha-Pinto} {et~al.}(2002){Rocha-Pinto}, {Castilho}, \&
  {Maciel}}]{rocha-pinto02}
{Rocha-Pinto}, H.~J., {Castilho}, B.~V., \& {Maciel}, W.~J. 2002, \aap, 384,
  912

\bibitem[{{Saffe} {et~al.}(2016){Saffe}, {Flores}, {Jaque Arancibia},
  {Buccino}, \& {Jofr{\'e}}}]{saffe16}
{Saffe}, C., {Flores}, M., {Jaque Arancibia}, M., {Buccino}, A., \&
  {Jofr{\'e}}, E. 2016, \aap, 588, A81

\bibitem[{{Salaris} {et~al.}(1993){Salaris}, {Chieffi}, \&
  {Straniero}}]{salaris93}
{Salaris}, M., {Chieffi}, A., \& {Straniero}, O. 1993, \apj, 414, 580

\bibitem[{{Santos} {et~al.}(2004){Santos}, {Israelian}, {Randich}, {Garc{\'\i}a
  L{\'o}pez}, \& {Rebolo}}]{santos04}
{Santos}, N.~C., {Israelian}, G., {Randich}, S., {Garc{\'\i}a L{\'o}pez},
  R.~J., \& {Rebolo}, R. 2004, \aap, 425, 1013

\bibitem[{{Schneider} {et~al.}(2012){Schneider}, {Le Sidaner}, {Savalle}, \&
  {Zolotukhin}}]{schneider12}
{Schneider}, J., {Le Sidaner}, P., {Savalle}, R., \& {Zolotukhin}, I. 2012,
  Astronomical Society of the Pacific Conference Series, Vol. 461, {The
  exoplanet.eu Database and Associated VO Services}, ed. P.~{Ballester},
  D.~{Egret}, \& N.~P.~F. {Lorente}, 447

\bibitem[{{Sch{\"o}nrich} \& {Binney}(2009)}]{schonrich09}
{Sch{\"o}nrich}, R. \& {Binney}, J. 2009, \mnras, 396, 203

\bibitem[{{Schuster} \& {Nissen}(1989)}]{schuster89}
{Schuster}, W.~J. \& {Nissen}, P.~E. 1989, \aap, 221, 65

\bibitem[{{Sellwood} \& {Binney}(2002)}]{sellwood02}
{Sellwood}, J.~A. \& {Binney}, J.~J. 2002, \mnras, 336, 785

\bibitem[{{Silva Aguirre} {et~al.}(2018){Silva Aguirre}, {Bojsen-Hansen},
  {Slumstrup}, {Casagrande}, {Kawata}, {Ciuc{\v{a}}}, {Hand berg}, {Lund},
  {Mosumgaard}, {Huber}, {Johnson}, {Pinsonneault}, {Serenelli}, {Stello},
  {Tayar}, {Bird}, {Cassisi}, {Hon}, {Martig}, {Nissen}, {Rix},
  {Sch{\"o}nrich}, {Sahlholdt}, {Trick}, \& {Yu}}]{silva-aguirre18}
{Silva Aguirre}, V., {Bojsen-Hansen}, M., {Slumstrup}, D., {et~al.} 2018,
  \mnras, 475, 5487

\bibitem[{{Silva Aguirre} {et~al.}(2015){Silva Aguirre}, {Davies}, {Basu},
  {Christensen-Dalsgaard}, {Creevey}, {Metcalfe}, {Bedding}, {Casagrande},
  {Handberg}, {Lund}, {Nissen}, {Chaplin}, {Huber}, {Serenelli}, {Stello}, {Van
  Eylen}, {Campante}, {Elsworth}, {Gilliland}, {Hekker}, {Karoff}, {Kawaler},
  {Kjeldsen}, \& {Lundkvist}}]{silva-aguirre15}
{Silva Aguirre}, V., {Davies}, G.~R., {Basu}, S., {et~al.} 2015, \mnras, 452,
  2127

\bibitem[{{Silva Aguirre} {et~al.}(2017){Silva Aguirre}, {Lund}, {Antia},
  {Ball}, {Basu}, {Christensen-Dalsgaard}, {Lebreton}, {Reese}, {Verma},
  {Casagrande}, {Justesen}, {Mosumgaard}, {Chaplin}, {Bedding}, {Davies},
  {Handberg}, {Houdek}, {Huber}, {Kjeldsen}, {Latham}, {White}, {Coelho},
  {Miglio}, \& {Rendle}}]{silva-aguirre17}
{Silva Aguirre}, V., {Lund}, M.~N., {Antia}, H.~M., {et~al.} 2017, \apj, 835,
  173

\bibitem[{{Snaith} {et~al.}(2015){Snaith}, {Haywood}, {Di Matteo}, {Lehnert},
  {Combes}, {Katz}, \& {G{\'o}mez}}]{snaith15}
{Snaith}, O., {Haywood}, M., {Di Matteo}, P., {et~al.} 2015, \aap, 578, A87

\bibitem[{{Sousa} {et~al.}(2008){Sousa}, {Santos}, {Mayor}, {Udry},
  {Casagrande}, {Israelian}, {Pepe}, {Queloz}, \& {Monteiro}}]{sousa08}
{Sousa}, S.~G., {Santos}, N.~C., {Mayor}, M., {et~al.} 2008, \aap, 487, 373

\bibitem[{{Spina} {et~al.}(2018){Spina}, {Mel{\'e}ndez}, {Karakas}, {dos
  Santos}, {Bedell}, {Asplund}, {Ram{\'\i}rez}, {Yong}, {Alves-Brito}, {Bean},
  \& {Dreizler}}]{spina18}
{Spina}, L., {Mel{\'e}ndez}, J., {Karakas}, A.~I., {et~al.} 2018, \mnras, 474,
  2580

\bibitem[{{Spina} {et~al.}(2016){Spina}, {Mel{\'e}ndez}, {Karakas},
  {Ram{\'{\i}}rez}, {Monroe}, {Asplund}, \& {Yong}}]{spina16}
{Spina}, L., {Mel{\'e}ndez}, J., {Karakas}, A.~I., {et~al.} 2016, \aap, 593,
  A125

\bibitem[{{Spina} {et~al.}(2020){Spina}, {Nordlander}, {Casey}, {Bedell},
  {D'Orazi}, {Mel{\'e}ndez}, {Karakas}, {Desidera}, {Baratella}, {Yana
  Galarza}, \& {Casali}}]{spina20}
{Spina}, L., {Nordlander}, T., {Casey}, A.~R., {et~al.} 2020, \apj, 895, 52

\bibitem[{{Spitoni} {et~al.}(2019){Spitoni}, {Silva Aguirre}, {Matteucci},
  {Calura}, \& {Grisoni}}]{spitoni19}
{Spitoni}, E., {Silva Aguirre}, V., {Matteucci}, F., {Calura}, F., \&
  {Grisoni}, V. 2019, \aap, 623, A60

\bibitem[{{Spitoni} {et~al.}(2020){Spitoni}, {Verma}, {Silva Aguirre}, \&
  {Calura}}]{spitoni20}
{Spitoni}, E., {Verma}, K., {Silva Aguirre}, V., \& {Calura}, F. 2020, \aap,
  635, A58

\bibitem[{{Stonkut{\.{e}}} {et~al.}(2020){Stonkut{\.{e}}}, {Chorniy},
  {Tautvai{\v{s}}ien{\.{e}}}, {Drazdauskas}, {Minkevi{\v{c}}i{\={u}}t{\.{e}}},
  {Mikolaitis}, {Kjeldsen}, {Essen}, {Pak{\v{s}}tien{\.{e}}}, \&
  {Bagdonas}}]{stonkute20}
{Stonkut{\.{e}}}, E., {Chorniy}, Y., {Tautvai{\v{s}}ien{\.{e}}}, G., {et~al.}
  2020, \aj, 159, 90

\bibitem[{{Str{\"o}mgren} {et~al.}(1982){Str{\"o}mgren}, {Gustafsson}, \&
  {Olsen}}]{stromgren82}
{Str{\"o}mgren}, B., {Gustafsson}, B., \& {Olsen}, E.~H. 1982, \pasp, 94, 5

\bibitem[{{Su{\'a}rez-Andr{\'e}s} {et~al.}(2018){Su{\'a}rez-Andr{\'e}s},
  {Israelian}, {Gonz{\'a}lez Hern{\'a}ndez}, {Adibekyan}, {Delgado Mena},
  {Santos}, \& {Sousa}}]{suarez-andres18}
{Su{\'a}rez-Andr{\'e}s}, L., {Israelian}, G., {Gonz{\'a}lez Hern{\'a}ndez},
  J.~I., {et~al.} 2018, \aap, 614, A84

\bibitem[{{Thoul} {et~al.}(1994){Thoul}, {Bahcall}, \& {Loeb}}]{thoul94}
{Thoul}, A.~A., {Bahcall}, J.~N., \& {Loeb}, A. 1994, \apj, 421, 828

\bibitem[{{Titarenko} {et~al.}(2019){Titarenko}, {Recio-Blanco}, {de Laverny},
  {Hayden}, \& {Guiglion}}]{titarenko19}
{Titarenko}, A., {Recio-Blanco}, A., {de Laverny}, P., {Hayden}, M., \&
  {Guiglion}, G. 2019, \aap, 622, A59

\bibitem[{{Trilling} {et~al.}(2008){Trilling}, {Bryden}, {Beichman}, {Rieke},
  {Su}, {Stansberry}, {Blaylock}, {Stapelfeldt}, {Beeman}, \&
  {Haller}}]{Trilling08}
{Trilling}, D.~E., {Bryden}, G., {Beichman}, C.~A., {et~al.} 2008, \apj, 674,
  1086

\bibitem[{{Tucci Maia} {et~al.}(2019){Tucci Maia}, {Mel{\'e}ndez},
  {Lorenzo-Oliveira}, {Spina}, \& {Jofr{\'e}}}]{tucci-maia19}
{Tucci Maia}, M., {Mel{\'e}ndez}, J., {Lorenzo-Oliveira}, D., {Spina}, L., \&
  {Jofr{\'e}}, P. 2019, \aap, 628, A126

\bibitem[{{Tucci Maia} {et~al.}(2016){Tucci Maia}, {Ram{\'{\i}}rez},
  {Mel{\'e}ndez}, {Bedell}, {Bean}, \& {Asplund}}]{tucci-maia16}
{Tucci Maia}, M., {Ram{\'{\i}}rez}, I., {Mel{\'e}ndez}, J., {et~al.} 2016,
  \aap, 590, A32

\bibitem[{{Turcotte} {et~al.}(1998){Turcotte}, {Richer}, {Michaud}, {Iglesias},
  \& {Rogers}}]{turcotte98}
{Turcotte}, S., {Richer}, J., {Michaud}, G., {Iglesias}, C.~A., \& {Rogers},
  F.~J. 1998, \apj, 504, 539

\bibitem[{{van Leeuwen}(2007)}]{vanleeuwen07}
{van Leeuwen}, F. 2007, \aap, 474, 653

\bibitem[{{Venn} {et~al.}(2004){Venn}, {Irwin}, {Shetrone}, {Tout}, {Hill}, \&
  {Tolstoy}}]{venn04}
{Venn}, K.~A., {Irwin}, M., {Shetrone}, M.~D., {et~al.} 2004, \aj, 128, 1177

\bibitem[{{Verma} {et~al.}(2019){Verma}, {Raodeo}, {Basu}, {Silva Aguirre},
  {Mazumdar}, {Mosumgaard}, {Lund}, \& {Ranadive}}]{verma19}
{Verma}, K., {Raodeo}, K., {Basu}, S., {et~al.} 2019, \mnras, 483, 4678

\bibitem[{{Vorontsov} {et~al.}(1991){Vorontsov}, {Baturin}, \&
  {Pamiatnykh}}]{vorontsov91}
{Vorontsov}, S.~V., {Baturin}, V.~A., \& {Pamiatnykh}, A.~A. 1991, \nat, 349,
  49

\bibitem[{{Weiss} \& {Schlattl}(2008)}]{weiss08}
{Weiss}, A. \& {Schlattl}, H. 2008, \apss, 316, 99

\bibitem[{{Yana Galarza} {et~al.}(2019){Yana Galarza}, {Mel{\'e}ndez},
  {Lorenzo-Oliveira}, {Valio}, {Reggiani}, {Carlos}, {Ponte}, {Spina},
  {Haywood}, \& {Gandolfi}}]{galarza19}
{Yana Galarza}, J., {Mel{\'e}ndez}, J., {Lorenzo-Oliveira}, D., {et~al.} 2019,
  \mnras, 490, L86

\bibitem[{{Yi} {et~al.}(2001){Yi}, {Demarque}, {Kim}, {Lee}, {Ree}, {Lejeune},
  \& {Barnes}}]{yi01}
{Yi}, S., {Demarque}, P., {Kim}, Y.-C., {et~al.} 2001, \apjs, 136, 417

\end{thebibliography}

\end{document}